\documentclass[fleqn,usenatbib]{mnras}

\usepackage{newtxtext,newtxmath}

\usepackage[T1]{fontenc}

\DeclareRobustCommand{\VAN}[3]{#2}
\let\VANthebibliography\thebibliography
\def\thebibliography{\DeclareRobustCommand{\VAN}[3]{##3}\VANthebibliography}

\usepackage{graphicx}	
\usepackage{amsmath}	
\usepackage{lscape}
\usepackage{caption}
\usepackage{subcaption}
\usepackage{makecell}
\usepackage{threeparttable}
\usepackage{hyperref}
\usepackage{tabularx}
\usepackage{booktabs}
\usepackage{url}
\usepackage{enumitem}
\usepackage{ulem}
\usepackage{threeparttable}
\usepackage{cleveref}
\usepackage{natbib}
\usepackage{placeins}
\usepackage{hyperref}
\usepackage{physics}

\title[Non-radial oscillations in newly born HS]{Non-radial oscillations in newly born compact star considering effects of phase transition}

\author[A. Kumar et al.]{
Anil Kumar,$^{1}$\thanks{E-mail: anil.1@iitj.ac.in}
Pratik Thakur,$^{1}$
and Monika Sinha $^{1}$
\\
$^{1}$Indian Institute of Technology Jodhpur, Jodhpur 342037, India
}

\date{Accepted XXX. Received YYY; in original form ZZZ}

\pubyear{2023}

\begin{document}
\label{firstpage}
\pagerange{\pageref{firstpage}--\pageref{lastpage}}
\maketitle

\begin{abstract}
The massive stars end their lives by supernova explosions leaving central compact objects that may evolve into neutron stars. Initially, after birth, the star remains hot and gradually cools down. We explore the matter and star properties during this initial stage of the compact stars considering the possibility of the appearance of deconfined quark matter in the core of the star. At the initial stage after the supernova explosion, the occurrence of non-radial oscillation in the newly born compact object is highly possible. Non-radial oscillations are an important source of GWs. There is a high chance for GWs from these oscillations, especially the nodeless fundamental (f-) mode to be detected by next-generation GW detectors. We study the evolution in frequencies of non-radial oscillation after birth considering phase transition and predicting the possible signature for different possibilities of theoretical compact star models.
\end{abstract}



\section{Introduction}\label{intro}
\begin{figure*} 
  \centering
  \includegraphics[width=0.8\textwidth]{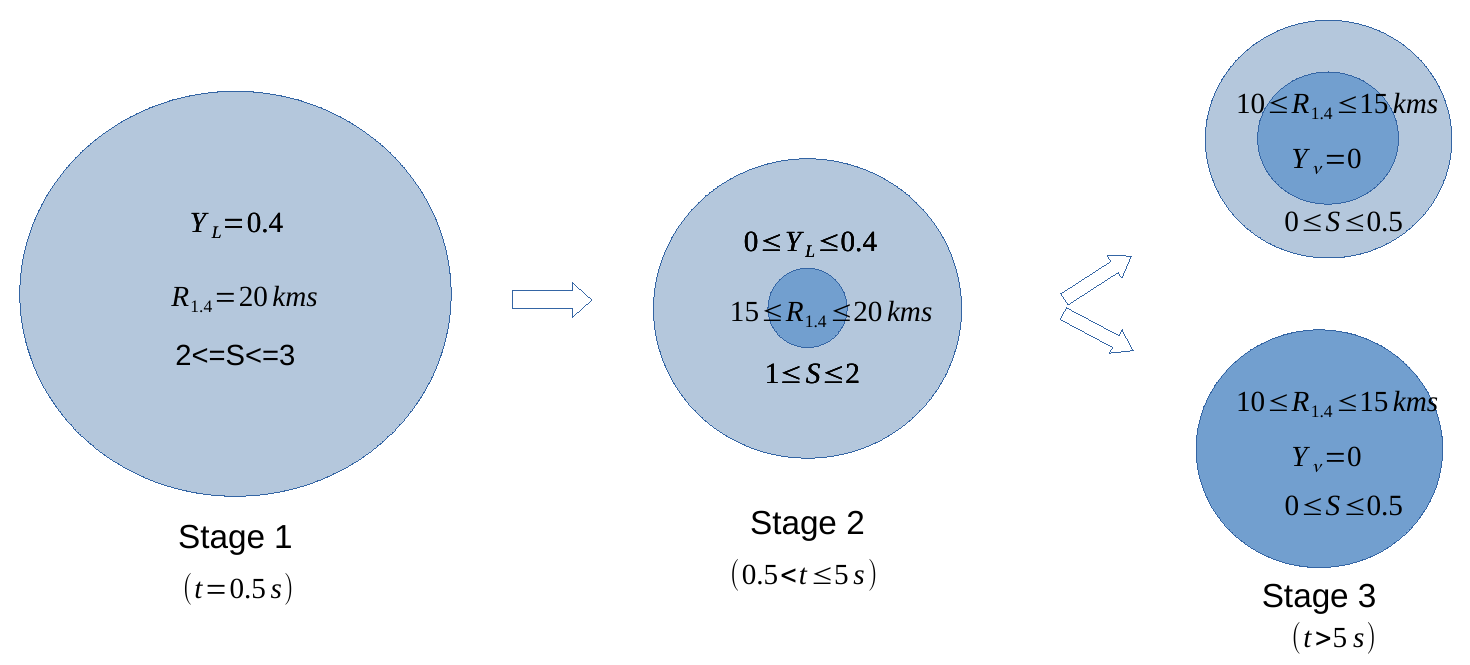}
  \caption{Possible stages of the star just after the supernova.}
  \label{fig:1a}
\end{figure*}
Stellar compact objects are the end products of the stellar evolution of massive stars. After the exhaustion of nuclear fuel, the star core lacks enough outward pressure to maintain the hydrostatic equilibrium and collapses under its gravity. This process leads to a supernova explosion and the collapsed core may become a neutron star, where the immense inward gravitational pull is balanced by the degeneracy pressure of constituent particles. At this stage, the remnant is highly dense with an average density around $10^{14}$ gm cm$^{-3}$ \citep{glendenning2012compact}. During the core collapse, the density of the remnant becomes so high that at first electrons are pushed into the nuclei. The electron Fermi energy is so high that in the inverse $\beta$-decay process, they combine with the protons and produce neutrons and electron neutrinos. After that, when the density crosses the neutron drip density the neutrons drip out of the nuclei. With further increase in density during the collapse, the nuclei merge with each other. Then after the supernova, the remnant contains the matter which is no longer composed of ions, but of mostly neutrons with a small admixture of protons and leptons. Then, within the remnant, due to the deleptonization, the matter pressure decreases, and consequently, the remnant contracts further. As a result, the density of matter further increases and gives a chance to exotic particles like hyperons or deconfined quarks to appear \citep{1988PThPh..80..861T,PhysRevLett.86.5223,2009PhRvL.102h1101S,glendenning2012compact,Lenka_2019,2019MNRAS.486.5441R,2021EPJA...57..270F,1993ApJ...414..701G,2011PhLB..704..343S,2021MNRAS.507.2991T}. Consequently, after the supernova, during the first a few seconds, inside the remnant core phase transition from nucleonic matter to deconfined quark matter may occur \citep{1999ApJ...513..780P,bauswein2023effects}.  
This deconfined quark matter, composed of up, down, and strange quarks, might be the true ground state of matter at high densities \citep{PhysRevD.4.1601,PhysRevD.30.272}. This leads to the possibility of a strange star (SS) formation that is entirely made of stable strange quark matter (SQM) up to the surface \citep{1986ApJ...310..261A}. If the entire star does not get converted to SS, it is possible that the deconfined quark core is surrounded by hadronic matter - the hybrid star (HS). 
Then the formation of HS as well as SS as a remnant of supernova are both possible depending on the underlying strong interaction nature inside highly dense matter. Until the true nature of strong interaction is well known, no single possibility can be ruled out theoretically. However, at this stage, the non-radial oscillation of the star matter is an important feature of the newborn star, and through the observation of non-radial modes of the newborn compact stars, the signature of one of these two possible scenarios can be detected. There are several quasi-normal modes (QNM) like the fundamental f-mode, pressure p-modes, gravity g-modes, spacetime w-mode, etc. \cite{1999LRR.....2....2K,lindblom1983quadrupole,detweiler1985nonradial}, each classified based on the restoring forces which work to bring the star to equilibrium. For example, the f- and p- modes, which are acoustic waves in the star, are restored by fluid pressure while g-modes which arise due to density discontinuities or temperature and composition variations are restored by gravity (buoyancy). Several of these modes can be excited during supernovae explosions \citep{2001MNRAS.320..307K,2011MNRAS.418..427S,2020PhRvD.101h4039V}. Quadrupolar oscillations (l=2) of all modes lead to the strongest emission of gravitational waves (GWs). Apart from the obvious reasons for viscosity inside the star, the energy emitted in GWs by the oscillations of neutron stars is so large that it is the main effect responsible for damping the vibration \citep{2008JPhCS.118a2005S}. This leads to a decay in the amplitude of the emitted GW with time. Since the detected GW is characterized both by the frequency and the damping time of the oscillation modes, a detailed investigation of both is necessary. Further, although the frequency may be detected with good accuracy, damping times of the waves may not \cite{2001MNRAS.320..307K}, leading to the need for theoretical estimates of the same. Thus, during the initial stages of compact stars, the evolution of quasi-normal modes is important and needs to be explored to understand the physics of supernovae and remnant \citep{2003MNRAS.342..629F,2016PhRvD..94d4043S}. 
Hence, our objective of this work is to study the evolution in non-radial frequencies during the formation of compact objects from their birth by supernova explosion which are expected to be observable from future GW detection. In the next section \ref{sec:non-rad}, we briefly describe the formalism to obtain the frequencies of non-radial oscillations expected to occur in newly born compact objects. Then we explore the possible composition of matter and properties of its constituents first and then the consequent structure and observable properties of newly born compact stars. In the next section \ref{sec:formalism} we discuss the evolution of matter properties and composition and consequently its effect on star structure and observable properties during the formation of a stable compact star after its birth as the supernova remnant. In the last section, we summarize the results and point out the distinct observable features of two theoretical possibilities of final stage compact objects.  

Throughout this work, we use the natural units with $c=\hbar=k_B=1$.
\begin{figure*}
 \centering
 \begin{subfigure}{0.49\textwidth}
  \centering
  \includegraphics[width=\linewidth]{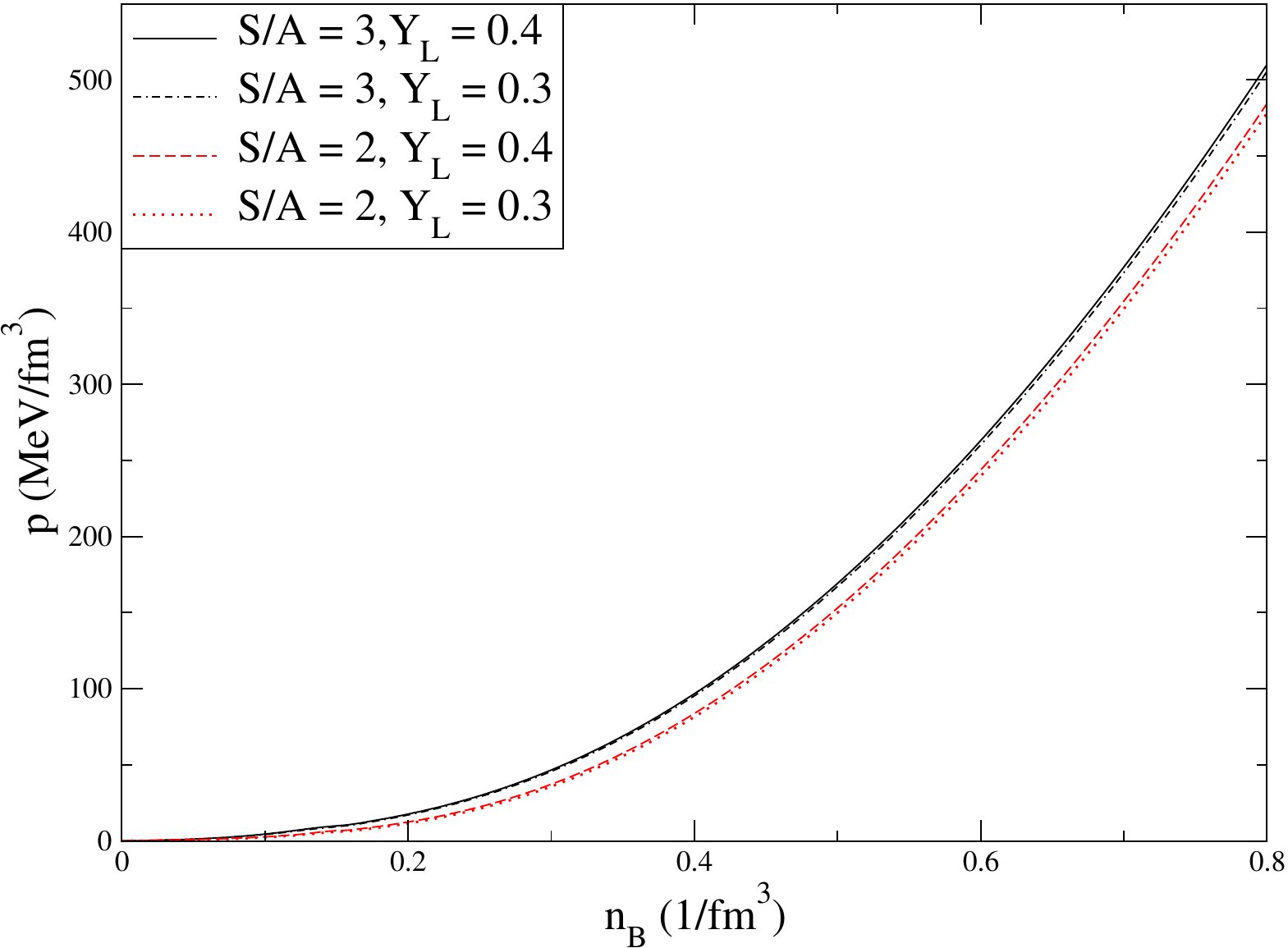}
 \end{subfigure}
 \centering
 \begin{subfigure}{0.49\textwidth}
  \includegraphics[width=\linewidth]{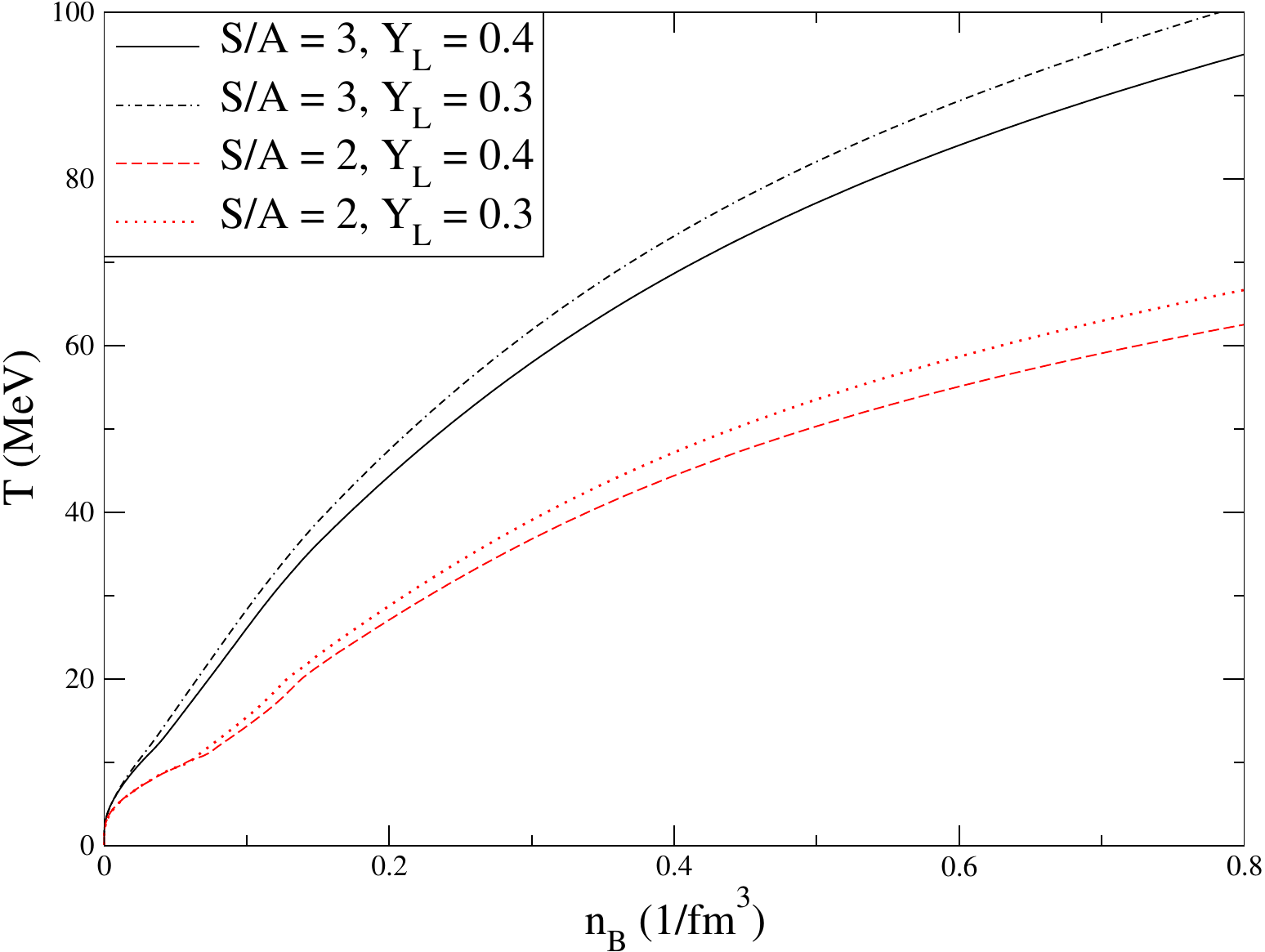}
 \end{subfigure}
 \caption{Left panel: Variation of matter pressure $p$, and  right panel: variation of temperature $T$ with baryonic number density $n_B$ for different $S/A$ and $Y_L$ in stage 1.}
\label{fig:eostemp1}
\end{figure*} 
\begin{figure*} 
\begin{center}
\includegraphics[width=9cm, keepaspectratio]{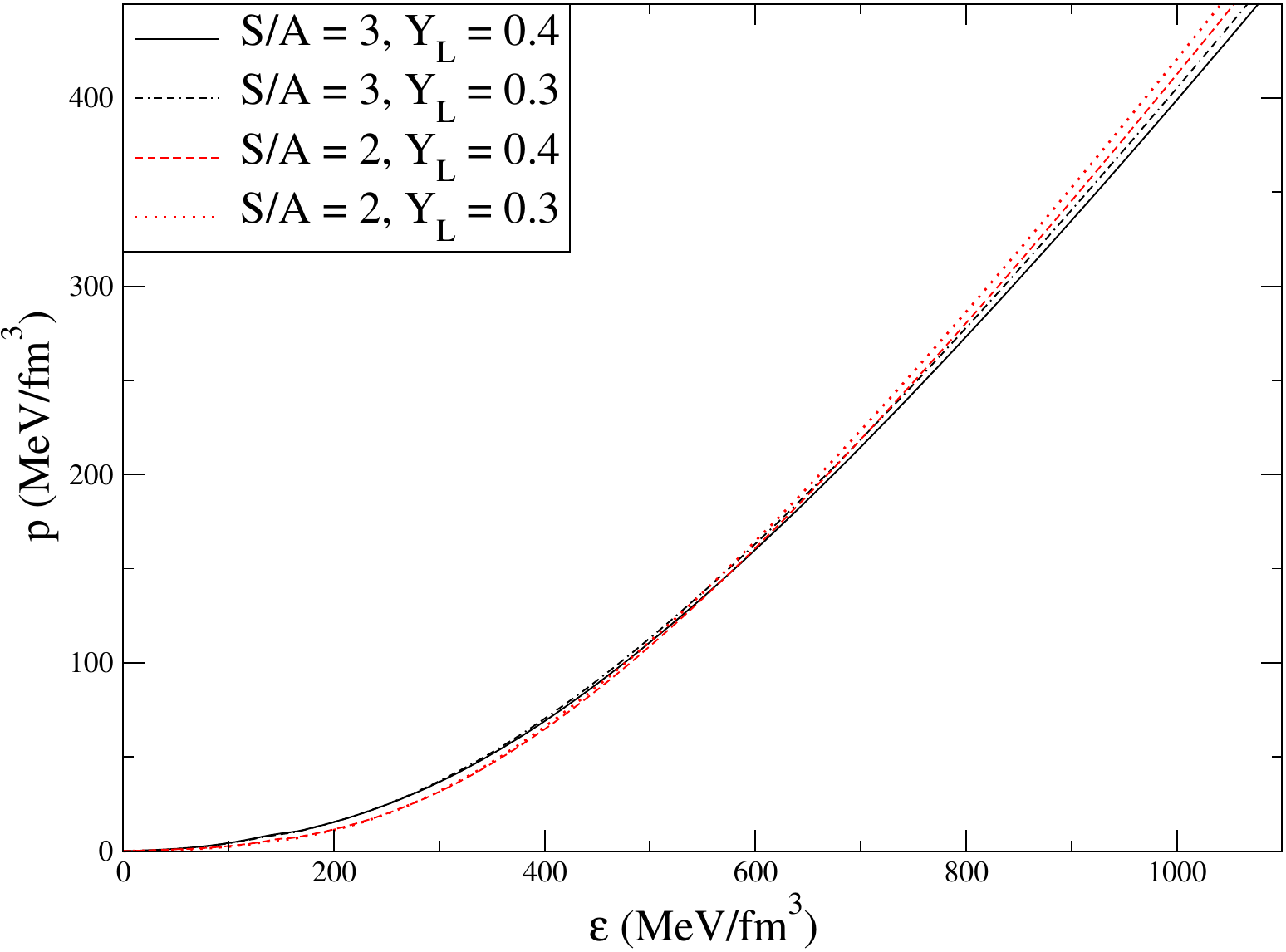}
\caption{Variation of pressure $p$ with energy density $\varepsilon$ for different $S/A$ and $Y_L$ in stage 1.}
\label{fig:eos1}
\end{center}
\end{figure*}
\section{Non-radial oscillations}\label{sec:non-rad}
Immediately, after the supernova explosion, the supernova remnant may be subjected to perturbation of stellar matter as well as of space-time metric which may lead to quasi-normal oscillations.
The theory of quasi-normal oscillations in relativistic stars was first given by the seminal work of Thorne and Campolattaro \citep{thorneNonRadialPulsationGeneralRelativistic1967}. Throughout this work, we adopt the Cowling classification, where the various modes are separated by the number of radial nodes \citep{coxNonradialOscillationsStars1976,rodriguezThreeApproachesClassification2023}. Our primary focus will be on the pressure driven modes among which f-mode, has no radial node number, while p-modes can have an arbitrary number of radial nodes greater than 0. 

QNMs arise due to perturbations of stellar matter and the spacetime metric, and an angular decomposition of these perturbations into spherical harmonics will contain even and odd parity components. The lowest frequency solution to the oscillation mode equation has no radial node and thus corresponds to the f-mode frequency ($\nu_f$). The next highest solution has one radial node and is thus the first p-mode (p1-mode). Although \cite{rodriguezThreeApproachesClassification2023} suggests that this classification fails before 0.4 seconds of the bounce, for the models in consideration in this work, the Cowling classification holds good.

In this work, we are interested in QNMs arising from fluid perturbations that couple to GWs, and thus restrict our focus to the dominant quadrupolar ($l=2$), even parity perturbations of the Regge-Wheeler metric \cite{thorneNonRadialPulsationGeneralRelativistic1967}:
\begin{multline}
    ds^2= -e^{2\Phi(r)}[1+r^lH_0(r)\mathcal{Y}_{lm}e^{i\omega t}]dt^2- 2i\omega r^{l+1}H_1(r)\mathcal{Y}_{lm}e^{i\omega t}dt dr\\+e^{2\Lambda(r)}[1-r^lH_0(r)\mathcal{Y}_{lm}e^{i\omega t}]dr^2\\+ r^2[1-r^lK(r)\mathcal{Y}_{lm}e^{i\omega t}][d\theta^2+ \sin^2\theta \ d\phi^2]
\end{multline}
Here $H_0$, $H_1$ and $K$ are the perturbation functions, $\mathcal{Y}_{lm}$ are the spherical harmonics and $\omega$ is the complex QNM frequency, the real part of which is the angular frequency of the mode, and the inverse of the positive imaginary part, the damping time. 

We employ full general relativistic calculations to determine $\nu_f$ which is discussed in \cref{sec: appendix f mode calculation}. The p-mode frequency ($\nu_p$) is exceptionally high, seemingly surpassing the detection range of future detectors. For the sake of completeness, we calculate $\nu_p$ using the Cowling approximation which is discussed in \cref{subsec: appendix Cowling approximation}.

\section{Evolution of newly born star}\label{sec:formalism} 

 During the supernova, a huge number of neutrinos are produced inside the core, whose mean free paths (in order of $100-1000$ m) depend on the temperature \citep{1999ApJ...513..780P,2002EPJA...13..383M,2012PhRvD..85h3003F}. After a few seconds of the explosion, a large number of neutrinos of all flavors are emitted leading to deleptonization. The diffusion of high energy electron neutrinos from core to surface generates a large amount of heat within the star raising the temperature in the range of $30-60$ MeV \citep{1999ApJ...513..780P, 1997PhR...280....1P,2012PhRvD..85h3003F,2022MNRAS.511..356P}. At the early stage ($t=0.5$ s), we assume the star is composed of lepton-rich nuclear matter at constant entropy. In the second stage ($t=0.5-5$ s), the nuclear matter may undergo a phase transition to quark matter inside the core. In this stage, the lepton fraction and entropy are lesser compared to the earlier stage due to the escape of highly energetic neutrinos \citep{1986ApJ...307..178B}. Finally, in the third stage, HS or SS is formed. If the remnant is not stable against the gravitational pull, it may get converted into a black hole. In this stage, the entropy of the star becomes almost constant, and neutrinos escape. This scenario is represented schematically in \cref{fig:1a}. We employ the standard evolution scenario to explore the possible states of compact stars and their matter constituents after a few seconds of the supernova. The observable properties of stars are highly correlated with the properties of matter. Consequently, theoretical models of matter inside newly generated compact stars are very much required \citep{2021Ap.....64..370H}. Here we discuss the possible matter composition and properties at this initial stage of the newly born star.
 
 During this time the star matter contains trapped neutrinos at high temperatures with constant entropy. Hence we discuss the lepton-rich matter at finite temperature and constant entropy - relevant for the newly born star interior. 
\subsection*{Stage 1}
Just immediately after the supernova, the matter is composed of mostly neutrons with a small admixture of protons and leptons (electrons and electron neutrinos). At this stage, 
entropy is high and matter is lepton-rich. The nucleonic matter at high density and finite temperature with constant entropy has been studied in several literature with several models \citep{2022MNRAS.517..507K,2021arXiv211005189F,2018PhRvC..98c5801B,2023arXiv230706892G,2023arXiv231019262G}. To study the matter at this stage,
we opt for the equation of state (EOS) HS(DD2) suggested by \citep{2010PhRvC..81a5803T}. In this work, the crust or lower density data for EOS has been taken from \textsc{CompOSE} database (\href{https://compose.obspm.fr/}{https://compose.obspm.fr/}). For nuclear matter EOS, we use density-dependent parametrization as given in \cref{sec: appendix EoS NM}. This model considers 
relativistic mean field (RMF) approach with 
density-dependent interaction. 
Within this model, nucleons interact among themselves via the three mesons: isoscalar-scalar $\sigma$, isoscalar-vector $\omega$, and isovector-vector $\rho$. 
The nuclear matter properties like nuclear saturation density ($n_{sat}$); energy per nucleon ($E/A$), skewness ($Q_{\text{sat}}$), compression modulus ($K_{\text{sat}}$)  at $n_{\text{sat}}$; symmetry energy ($J_{\text{sym}}$), its slope ($L_{\text{sym}}$), curvature ($K_{\text{sym}}$) at $n_{\text{sat}}$; Dirac effective mass ($m_{\text{eff}}$) for this model are given in \cref{table1}.
\begin{figure*}
 \centering
 \begin{subfigure}{0.49\textwidth}
  \centering
  \includegraphics[width=\linewidth]{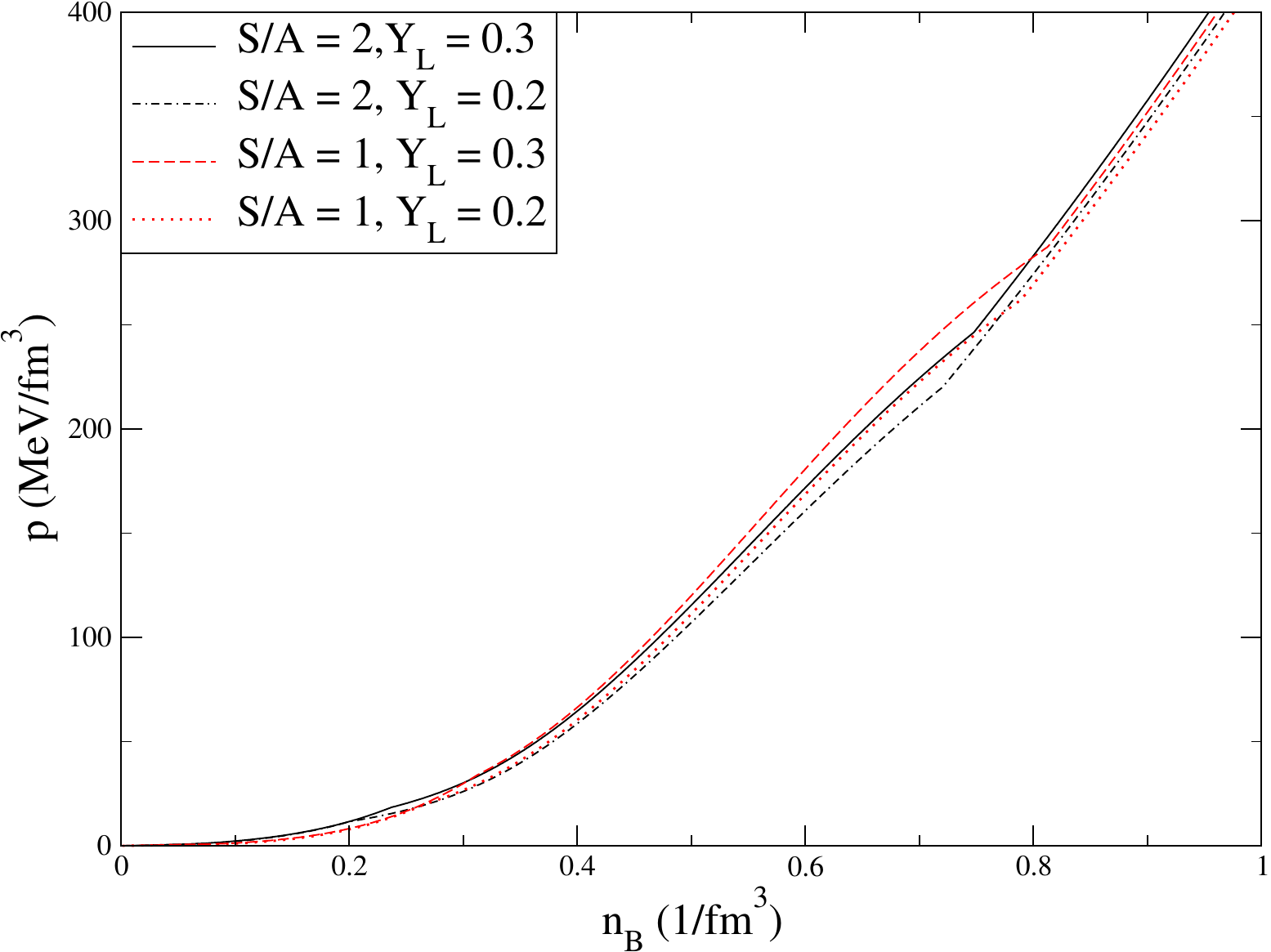}
 \end{subfigure}
 \centering
 \begin{subfigure}{0.49\textwidth}
  \includegraphics[width=\linewidth]{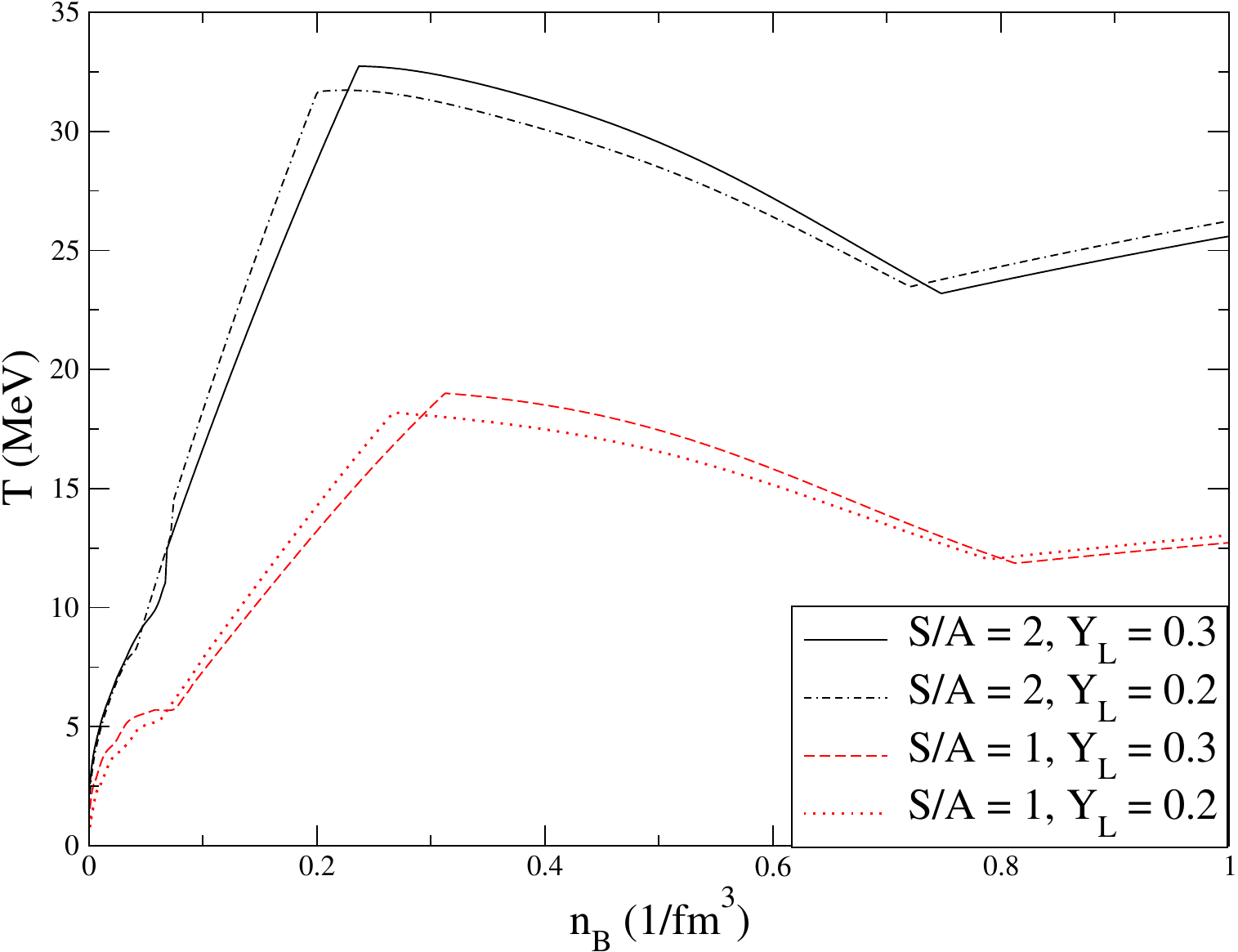}
 \end{subfigure}
 \caption{Variation of matter pressure $p$, and  right panel: variation of temperature $T$ with baryonic number density $n_B$ for different $S/A$ and $Y_L$ in stage 2.}
\label{fig:eostemp2}
\end{figure*}

\begin{figure*} 
\begin{center}
\includegraphics[width=9cm, keepaspectratio]{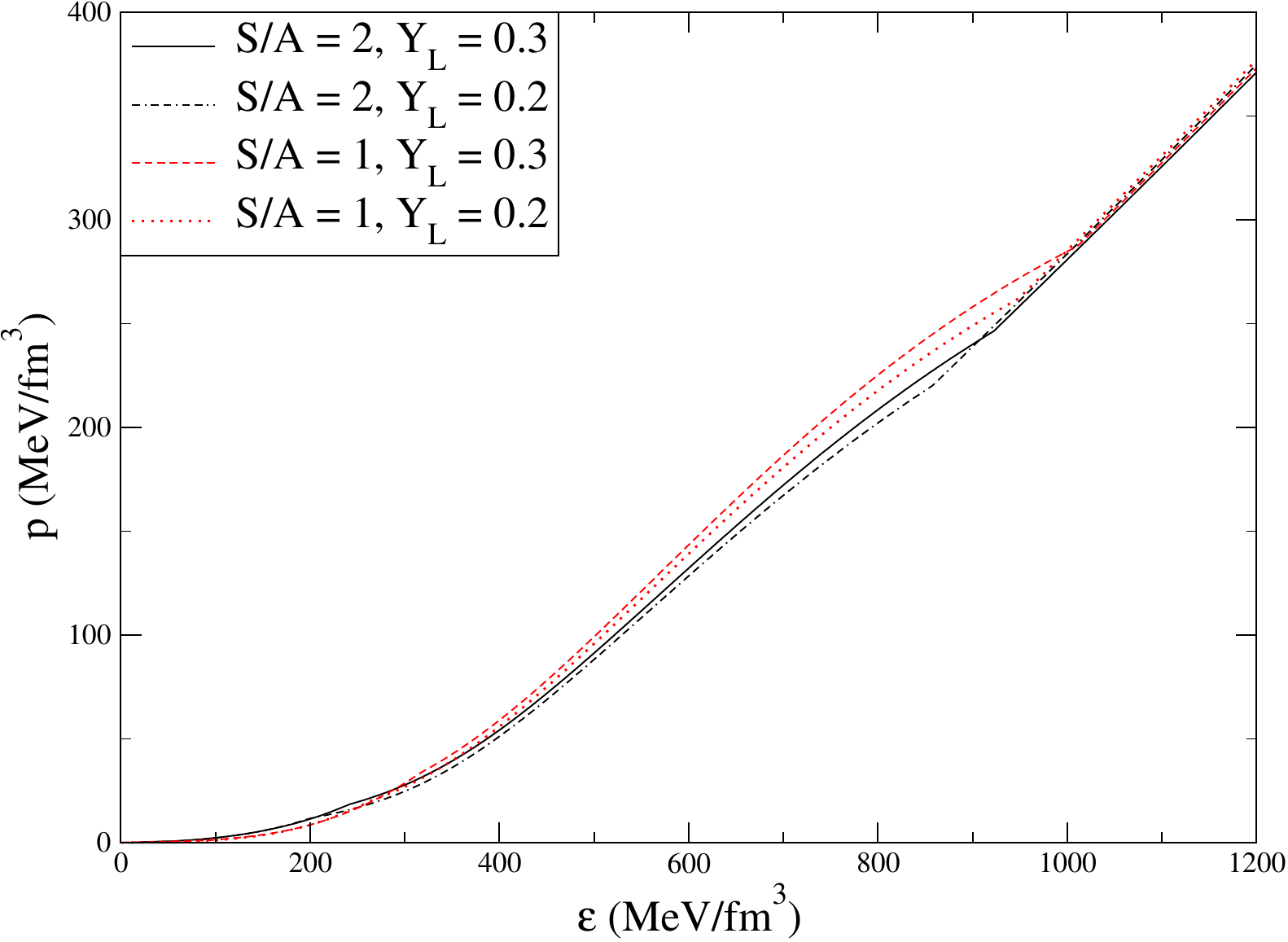}
\caption{Variation of pressure $p$ with energy density $\varepsilon$ for different $S/A$ and $Y_L$ in stage 2.}
\label{fig:eos2}
\end{center}
\end{figure*}
\begin{table*}
\begin{center}
\caption{Properties of nuclear matter with HS(DD2) EOS.}
\begin{tabular}{ccccccccccccc}
\hline \hline
       & $n_{\text{sat}}$ & $E/A$ & $Q_{\text{sat}}$& $K_{\text{sat}}$ & $J_{\text{sym}}$ &$L_{\text{sym}}$ & $K_{\text{sym}}$ & $m_{\text{eff}}$\\
     &($\text{fm}^{-3}$)& (MeV) & (MeV)& (MeV) & (MeV) & (MeV) & (MeV) & ($m_n$) \\         
\hline  & 0.149 & -16.02 & 168.65 & 242.72 & 31.67 & 55.04 & -93.23 & 0.563 &  \\
\hline
\end{tabular}
\label{table1}
\end{center} 
\end{table*}
The matter EOS, the variation of matter pressure $p$ with the energy density $\varepsilon$ and baryonic number density $n_B$, depends on the lepton fraction $Y_L = \frac{n_e+n_{\nu_e}}{n_B}$ with $n_e$ the number density of electrons and the entropy per baryon $S/A$. Hence, we use these quantities as parameters to determine the matter EoS. As in this stage entropy and lepton fraction are high we choose $S/A$ in the higher range of $2-3$ and $Y_L$ in the range of $0.3-0.4$. 

As a result temperature of a star is higher at this stage. For constant $S/A$ the temperature varies with density. The variation of $p$ and temperature $T$ with $n_B$ are shown in the left and right panels of \cref{fig:eostemp1} respectively. Naturally, $p$ increases for higher values of  $S/A$ and $Y_L$ as is evident from the figure. However, $p$ varies a little bit more with a variation of $S/A$ compared to the variation of $Y_L$. 
On the other hand, $T$ profile significantly depends on $S/A$ and $Y_L$. Naturally, $T$ is less for the lower value of $S/A$. For a particular value of $S/A$, $T$ is more for less value of $Y_L$. 
At a lower density regime, below the density range $200-300$ MeV fm$^{-3}$, the stiffness of EOS does not change significantly with the variation of $Y_L$. However, at higher density regime matter becomes a little stiff with lesser values of $Y_L$. With a decrease of $S/A$, the matter becomes soft but becomes stiff beyond density $450-550$ MeV fm$^{-3}$. These features are evident from \cref{fig:eos1}.

\subsection*{Stage 2}
While transitioning from stage 1 to stage 2, deleptonization occurs rapidly which causes a reduction in the pressure of the matter reducing the outward balancing force against gravity. The size of the star decreases and its density becomes higher. At such a high density of matter, there is the possibility of the presence of deconfined quark matter inside the core of the star. The matter properties with the possibility of deconfined quark appearance at finite temperature have been discussed in several previous literatures \citep{2011PhLB..704..343S,2018PhRvC..98e5805R,2000PhLB..486..239S}. Considering this possibility, we construct the matter using DD2 EOS for nuclear matter and vBAG model for quark matter. For SQM, the simplest model is the MIT Bag model; wherein quarks are assumed to be confined within a Bag and they are separated from the vacuum by the Bag pressure $B$. In the preliminary model, quarks are assumed to be free within the Bag. However, the maximum attainable mass of SSs in this model does not approach the observed lower limit of the maximum mass of compact objects. The inter-quark interaction term has been introduced within of Bag model in many works \citep{PhysRevD.30.2379,2001PhRvD..63l1702F,2005ApJ...629..969A}. One way to include the inter-quark interaction is via a vector particle. This is popularly known as the vBAG model. With this model, the SSs satisfy all the observed mass-radius relations and the limit on tidal deformability from GW observations. For a detailed discussion, readers may refer to \citep{2022MNRAS.513.3788K,PhysRevD.107.063024}. We construct the SQM at non-zero temperature considering the vector bag (vBAG) model. A detailed discussion of the calculation of SQM EOS is given in \cref{sec: appendix EoS SQM}. At finite temperature, equilibrium between the nuclear and quark phases can be established through Maxwell construction \cite{2022EPJA...58..152I}, Gibbs construction \cite{2000PhLB..486..239S,2022EPJA...58...55L,2013A&A...551A..13C}, and hadron-quark crossover \citep{2016PTEP.2016b1D01M}. During the transition to quark matter in neutron stars, the finite-size effects may change the geometrical structure and size of mix phase \citep{PhysRevLett.70.1355,2014PhRvC..89f5803Y,2017PhRvC..96b5802W}. We use Gibbs construction for the phase transition, which means there exists a mixed phase (MP) region between the hadronic and quark phases.  
In this MP region chemical equilibrium ($\mu_H=\mu_Q$), mechanical equilibrium ($P_H=P_Q$), and thermal equilibrium ($T_H=T_Q$) are maintained, where $H$ and $Q$ suffixes denote the hadronic and quark phases respectively. We opt for higher values of the bag parameter ($B^{1/4}=162$ MeV) than the maximum value of $B$ from the stability window of stable SQM to avoid the possibility of formation of a SS in stage 2. The maximum values of $B$ for stable SQM are given in \cref{tab:Bmax_stage2}.

\begin{table}
\centering
\caption{Maximum values of $B$ for stable SQM in stage 2}
\begin{tabular}{|l|l|l|l|}
\hline
\begin{tabular}[c]{@{}l@{}}S/A\\ ($k_B$)\end{tabular} & \begin{tabular}[c]{@{}l@{}}$G_V$\\ (fm$^2$)\end{tabular} & $Y_L$ & \begin{tabular}[c]{@{}l@{}}$B_\text{max}^{1/4}$\\ (MeV)\end{tabular} \\ \hline
2                                                  & 0.2                                                  & 0.3  & 140                                                  \\ \hline
2                                                  &0.2                                                  & 0.2  & 143                                                  \\ \hline
1                                                  & 0.2                                                  & 0.3  & 141.5                                                \\ \hline
1                                                  & 0.2                                                  & 0.2  & 144.7                                                \\ \hline
\end{tabular}
\label{tab:Bmax_stage2}
\end{table}
According to this matter composition, pressure $p$ with $n_B$ is shown in the left panel of \cref{fig:eostemp2} for different values of $S/A$ and $Y_L$. In this stage, the $Y_L$ is less compared to stage 1, and hence we consider $Y_L$ in the range $0.2-0.3$. Due to thermal energy loss, in this stage $S/A$ also reduces. We choose $S/A$ in the range of $1-2$. Naturally, for the higher $S/A$ the quarks appear earlier compared to the lower $S/A$. Consequently, the MP starts and ends earlier for higher $S/A$. As the appearance of quarks reduces $p$, the early appearance of quarks for higher $S/A$ makes lower $p$ for higher $S/A$ in the MP. After the end of the MP, in the pure quark phase, $p$ is higher for higher $S/A$ as expected.
 In the quark phase, $p$ is affected more by $Y_L$ as compared to the value of $S/A$. In the quark phase,  $p$ is higher for higher $Y_L$. Here also we notice that lower values of $Y_L$ make the phase transition earlier. In the temperature profile, as shown in the right panel of the same figure, the phase transition feature is evident. Temperature decreases in the MP region due to the consumption of heat in the form of latent heat. Then temperature increases in the quark phase. The rate of increase of temperature is less in the quark phase as compared to the hadronic phase. 

As shown in \cref{fig:eos2}, at a lower density regime as the matter is composed of only hadronic matter the EOS is softer for less $S/A$ as expected from the stage 1 result. The appearance of quarks softens the EOS. As higher $S/A$ makes the earlier phase transition to the MP and hence to the pure quark phase, the matter becomes softer for higher $S/A$ in the MP as well as in the pure quark phase. Pressure in the MP region is a combined effect of both phases, matter is softer for less $Y_L$. However, as in the case of nuclear matter, in the quark phase, the EOS of matter becomes stiffer at lower $Y_L$. In this phase, $Y_L$ affects more as compared to $S/A$.

\subsection*{Stage 3}
\begin{figure*} 
\begin{center}
\includegraphics[width=9cm, keepaspectratio]{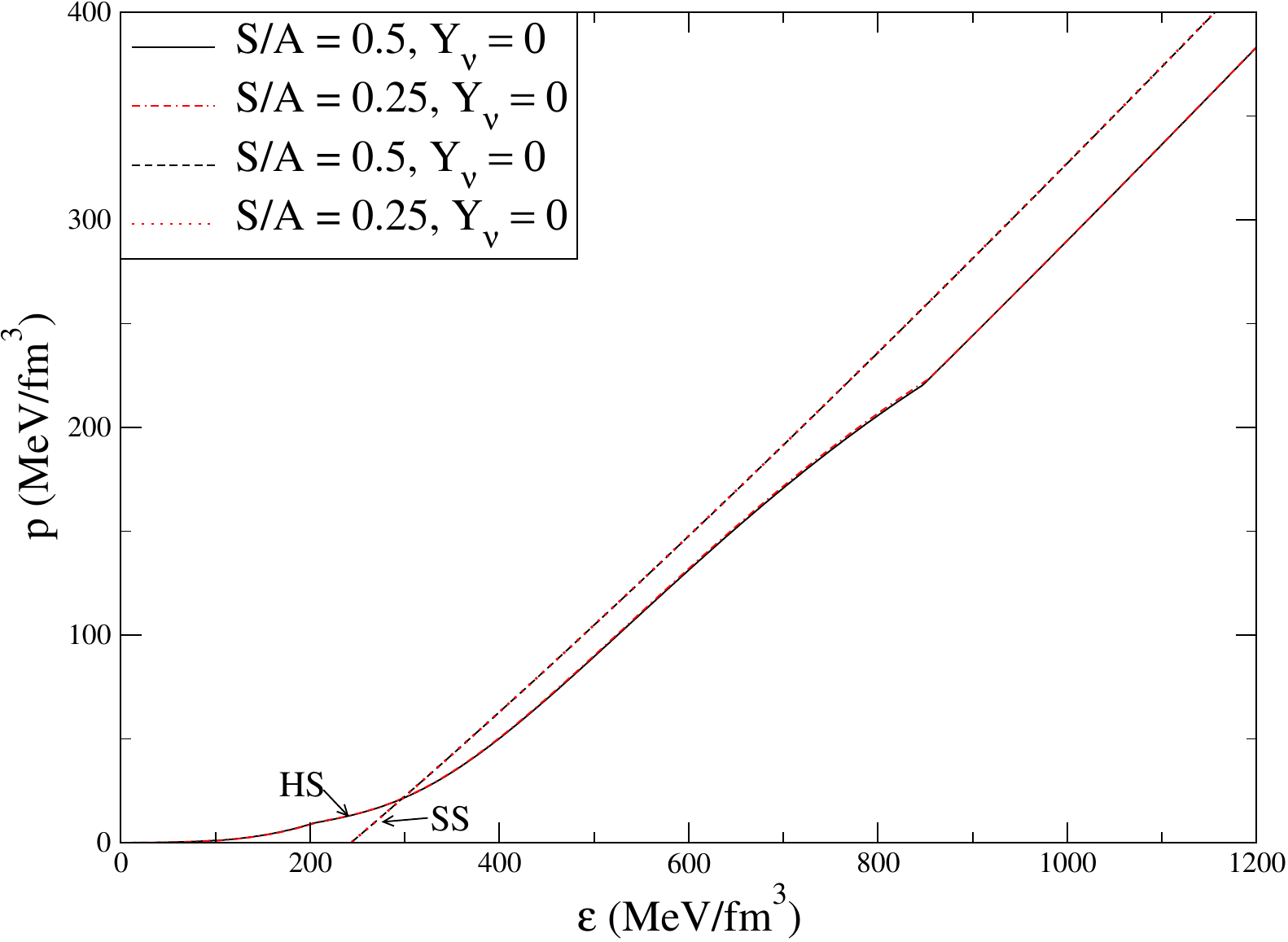}
\caption{Variation of matter pressure $p$ with energy denisty $\varepsilon$ for different $S/A$ and $Y_L$ in stage 3 for both HS and SS.}
\label{fig:eos3}
\end{center}
\end{figure*}

At stage 3, after deleptonization ($\mu_\nu=0$), the temperature reduces further and the remnant gets stabilized. As in this stage, the temperature gets reduced, the star contracts further, and the density of matter inside the star is enhanced. Consequently, in this stage, a phase transition occurs at a smaller temperature as compared to stage 2. Hence, as already mentioned earlier, in this final stage of the compact star after a supernova, there are two possibilities - the star may be HS with SQM at the core of the star surrounded by hadronic matter or SS entirely composed of SQM up to the surface without any crust. For HS, similar to stage 2, we use higher values of $B^{1/4}=162$ MeV than the maximum values of $B$ from stable SQM. The maximum values of $B$ for stable SQM are given in \cref{tab:Bmax_stage3}.
\begin{table}
\centering
\caption{Maximum values of $B$ for stable SQM in stage 3.}
\begin{tabular}{|l|l|l|}
\hline
\begin{tabular}[c]{@{}l@{}}S/A\\ ($k_B$)\end{tabular} & \begin{tabular}[c]{@{}l@{}}$G_V$\\ (fm$^2$)\end{tabular} & \begin{tabular}[c]{@{}l@{}}$B_\text{max}^{1/4}$\\ (MeV)\end{tabular} \\ \hline
0.5                                                 & 0.2                                                 & 149.5                                                  \\ \hline
0.25                                                 & 0.2                                                  & 149.8                                                  \\ \hline
\end{tabular}
\label{tab:Bmax_stage3}
\end{table}
In stage 3, we are also considering the possibility of the formation of SS. In that case, we choose a lower value of $B^{1/4}$ as $149$ MeV for SSs.
\begin{figure*}
 \centering
 \begin{subfigure}{0.49\textwidth}
  \centering
  \includegraphics[width=\linewidth]{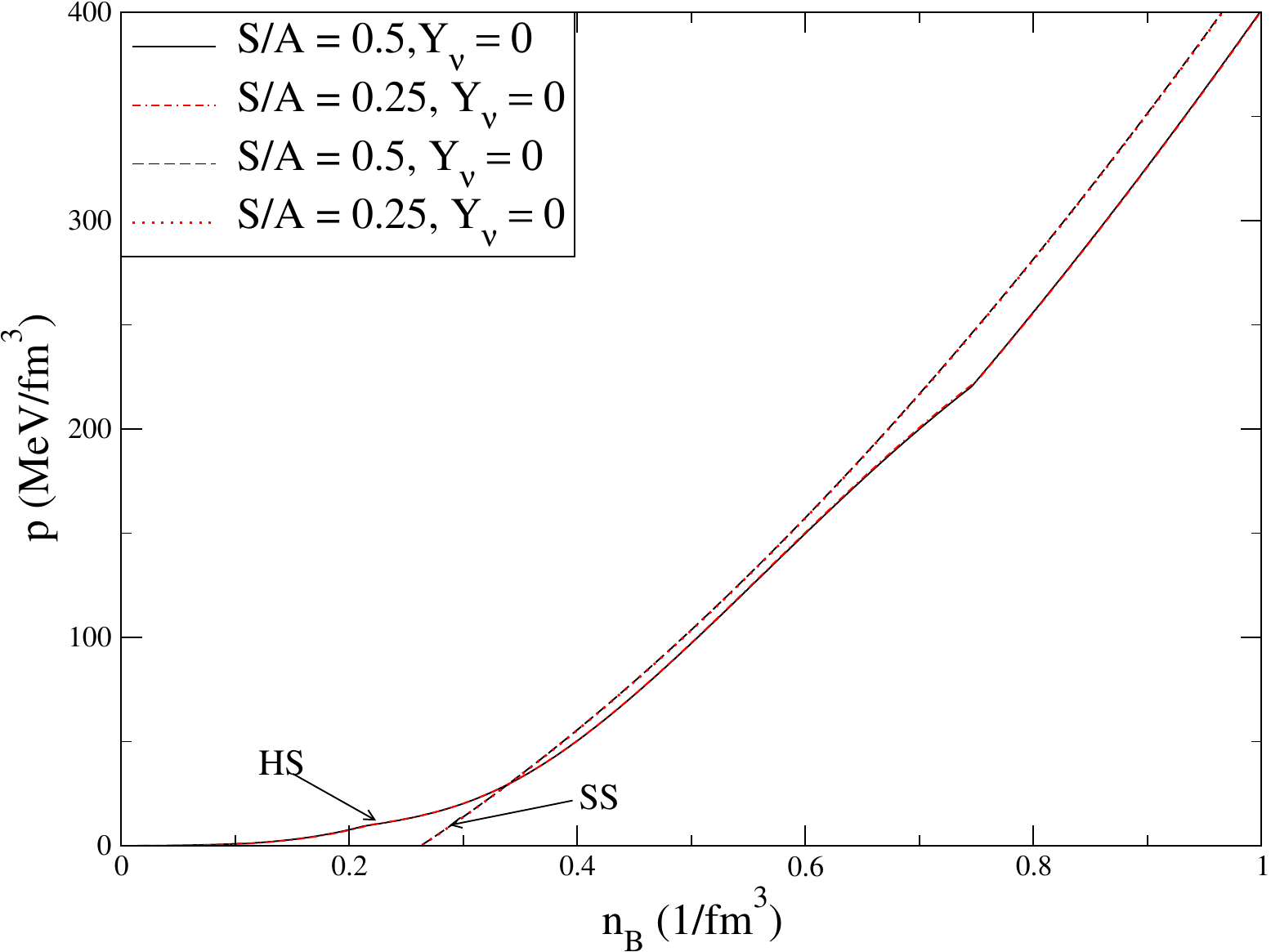}
 \end{subfigure}
 \centering
 \begin{subfigure}{0.49\textwidth}
  \includegraphics[width=\linewidth]{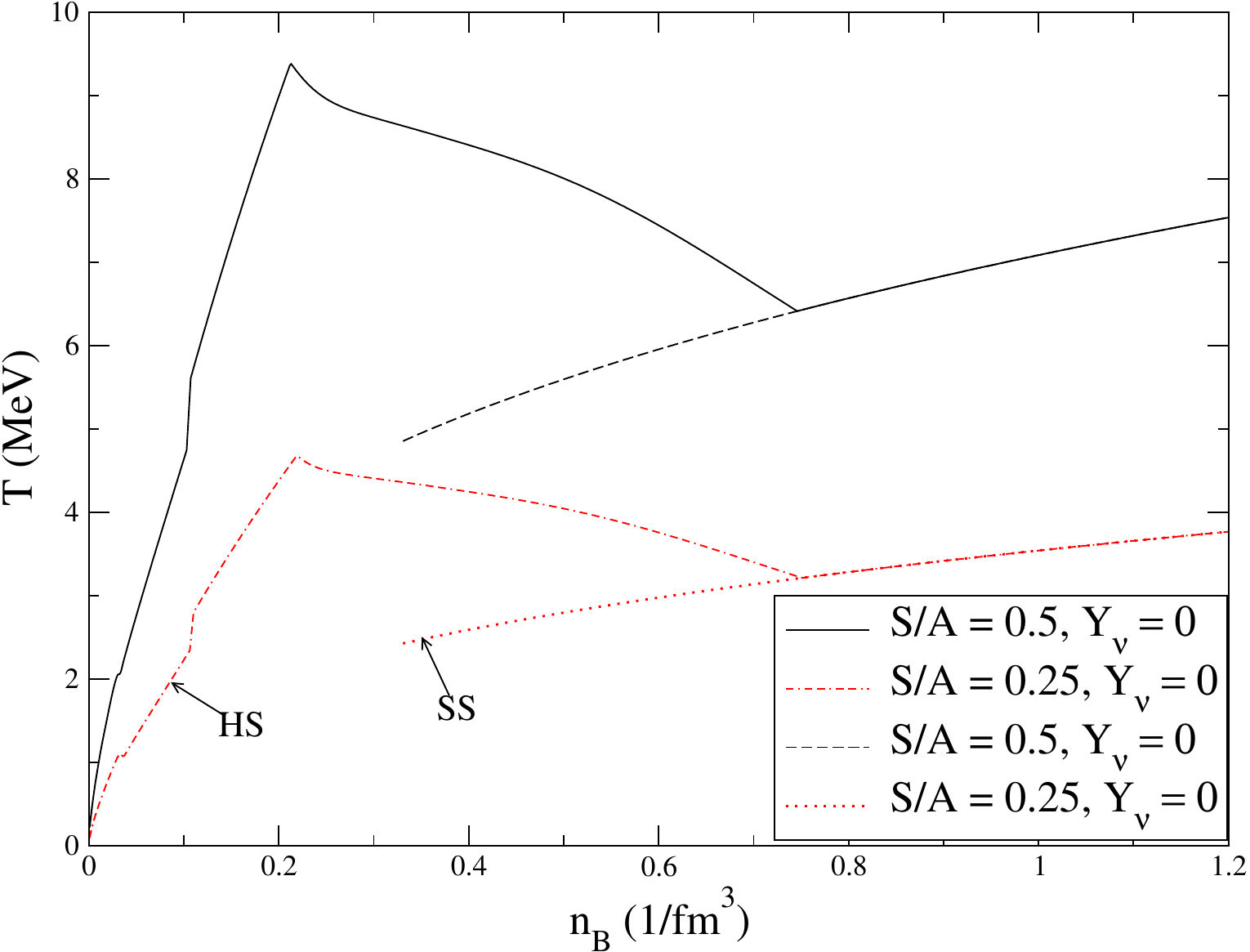}
 \end{subfigure}
 \caption{Variation of matter pressure $p$, and  right panel: variation of temperature $T$ with baryonic number density $n_B$ for different $S/A$ and $Y_L$ in stage 3 for both HS and SS.}
\label{fig:eostemp3}
\end{figure*}
Similar to the previous stages, we represent the variation of $p$ and $T$ with $n_B$ at the final stage in \cref{fig:eostemp3} for both possibilities. As less $Y_L$ makes the phase transition to quark matter earlier, in stage 3 deleptonization makes the transition to quark matter near density $n_B = 1.4~n_{sat}$. Hence there exists the possibility that the entire star is composed of SQM depending on the exact properties of matter at high density. It can be noticed that the temperature is lower at this stage compared to previous stages. The change in $p$ is very small with $S/A$ but there is a noticeable change in $T$ with $S/A$. At this stage, the EOS is represented in Fig. \ref{fig:eos3}. For SS, EOS becomes stiffer than HS because a lower value of $B$ is chosen.

\subsection{Properties of the star at different stages}   
\begin{figure*} 
\begin{center}
\includegraphics[width=15cm, keepaspectratio]{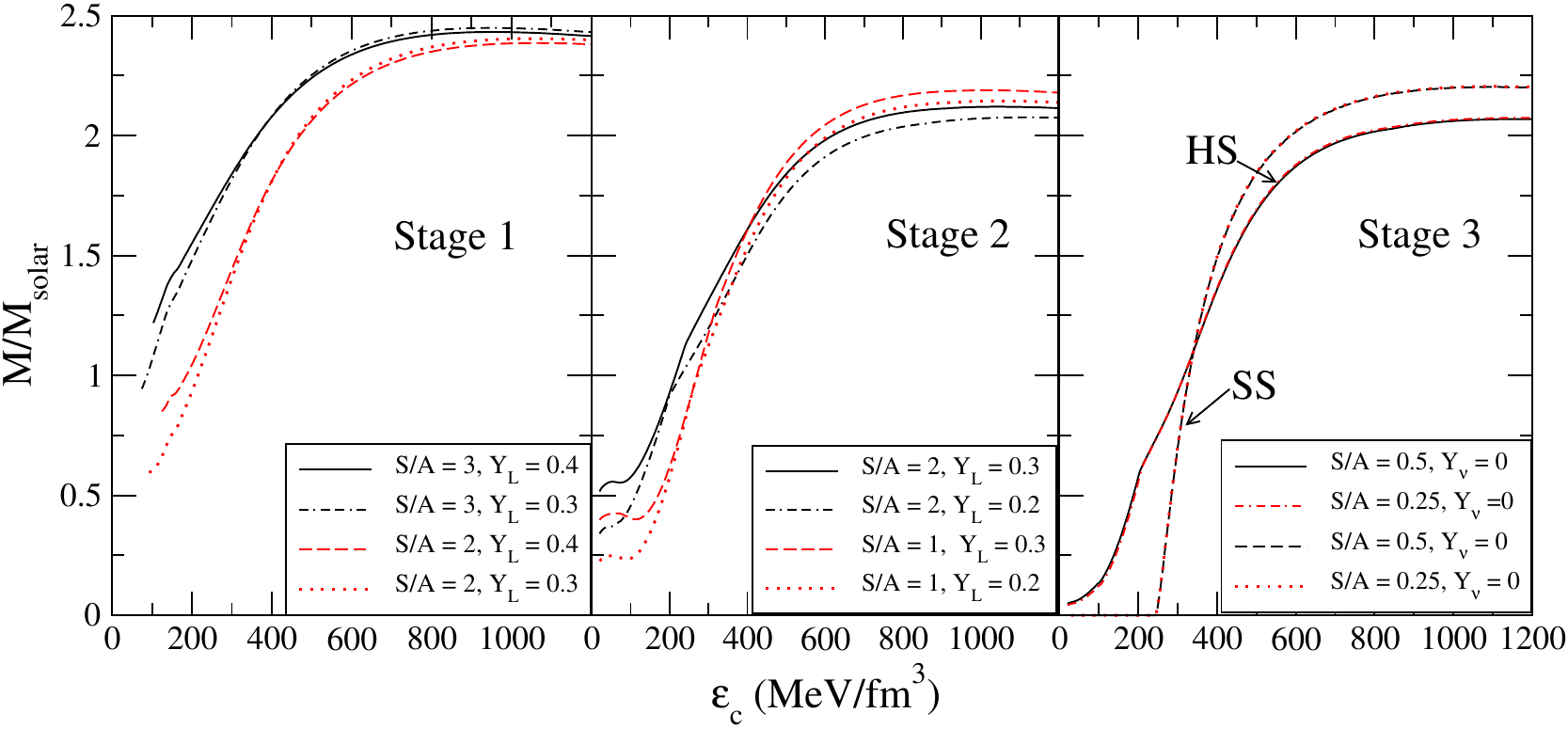}
\caption{Variation of star mass $M$ in unit of $M_\odot$ with central energy density $\varepsilon_c$ at different stages for different values of $S/A$ and $Y_L$.}
\label{fig:MvsCED}
\end{center}
\end{figure*}

\begin{figure*} 
\begin{center}
\includegraphics[width=15cm, keepaspectratio]{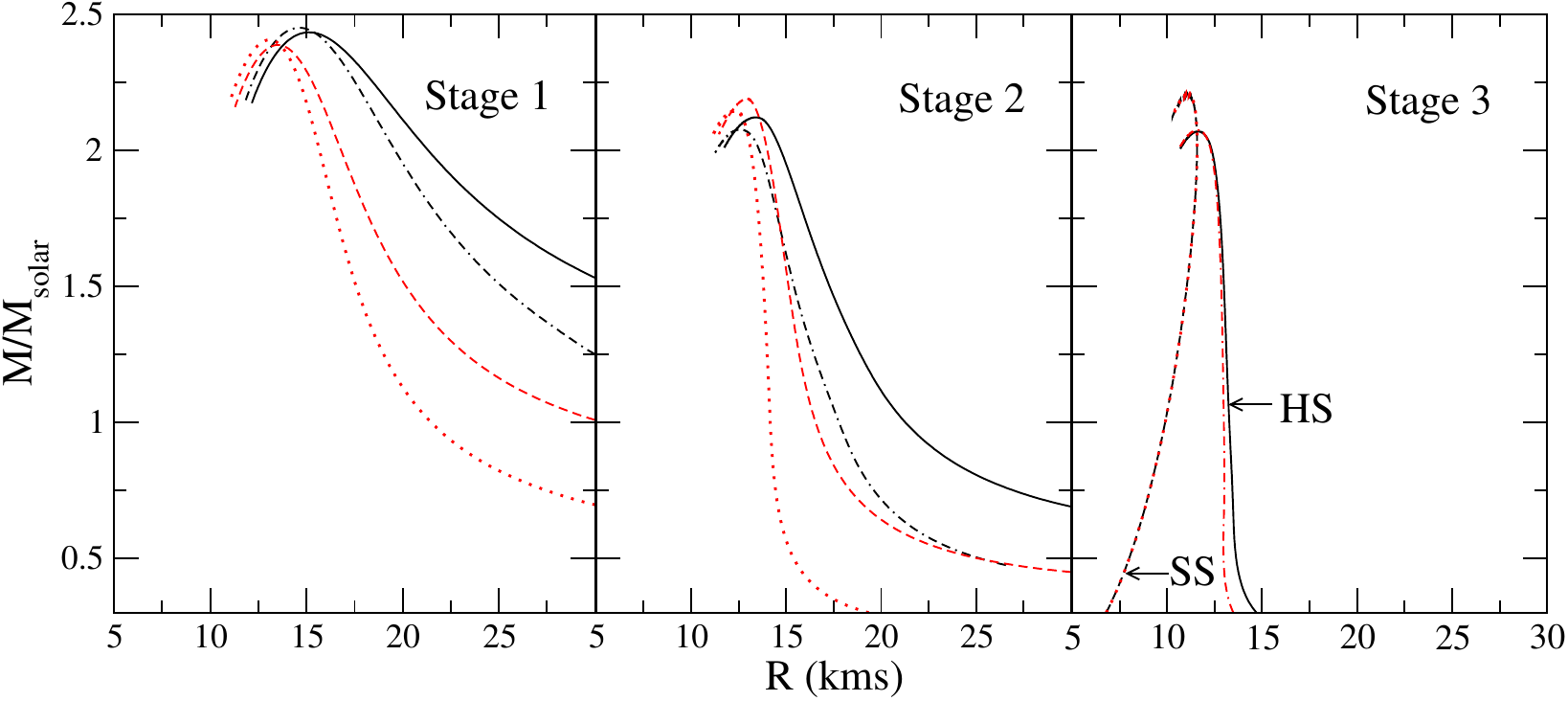}
\caption{M-R relations at different stages for different values of $S/A$ and $Y_L$. All legends are similar to fig. \ref{fig:MvsCED}.}
\label{fig:2}
\end{center}
\end{figure*}

As already discussed, the supernova remnant becomes more and more compact gradually during the evolution from stage 1 to stage 3 due to deleptonization and finally stabilizes. With the matter properties during these stages as discussed above, the star structure is obtained 
by solving the Tolman-Oppenheimer-Volkoff (TOV) equations. Here we discuss the star and properties of matter in the course of the evolution in a few seconds after the supernova to form the final stellar remnants.

For different central energy densities $\varepsilon_c$, the variation of stellar mass $M$ of stars is shown in fig. \ref{fig:MvsCED}. 
As expected, the $M$ increases with central energy density and achieves a maximum value at a specific central density. 
In stage 1, for a particular value of $\varepsilon_c$ mass decreases with the decrease of $S/A$ as in the range of matter density prevailing in most parts of the star, the matter becomes softer with the low value of $S/A$. In stage 2, when $\varepsilon_c$ is less, the star is dominated by pure nuclear matter, and the same feature is observed. However, in this stage for higher $\varepsilon_c$, the mass increases with $S/A$ as matter becomes softer for higher $S/A$ due to early phase transition to MP. In stage 3, for HS as the star is dominated by MP, the mass increases with the decrease of $S/A$ for HS. A similar feature is observed for SS.  Now central energy density corresponding to the maximum mass of the star is larger compared to stage 1. As a result of softer EOS due to the appearance of quark matter, we obtain the smaller maximum mass. However, the stellar at a particular $\varepsilon_c$ does not change significantly with $Y_L$. Still for low $\varepsilon_c$, mass decreases and in the higher $\varepsilon_c$ regime it increases with the decrease of $Y_L$ as expected from matter behavior. Also, with decreasing value of $Y_L$, the obtainable maximum mass increases in stage 1 as matter becomes stiffer with less $Y_L$ at high density. However, in stage 2, as the star is dominated by matter in MP, with decreasing $Y_L$ maximum obtainable mass decreases. As the decrease in $Y_L$ makes quark appearance earlier softening the matter, the mass decreases with decrease of $Y_L$ in this stage.
The mass-radius (M-R) relations during these stages are shown in \cref{fig:2}. 
In stage 1, large $S/A$ and $Y_L$ tend to increase the pressure. As a result of that EOS becomes stiff and the radius becomes very large for the same mass of star. At stage 2 when deleptonization occurs and entropy decreases the EOS becomes comparatively softer. This reduces the radius and increases the density of matter inside the star. This leads to the appearance of quark matter which softens the matter further. Hence from stage 1 to stage 3, the star becomes more compact which is evident from fig. \ref{fig:2}. At stage 3, when phase transition occurs early resulting larger quark core inside the star, the radius further reduces. The star becomes very compact and stable at the end. The maximum attainable mass in different stages with their central energy density and radius are tabulated in table \ref{tab:1}.

\begin{figure*} 
\begin{center}
\includegraphics[width=15cm, keepaspectratio]{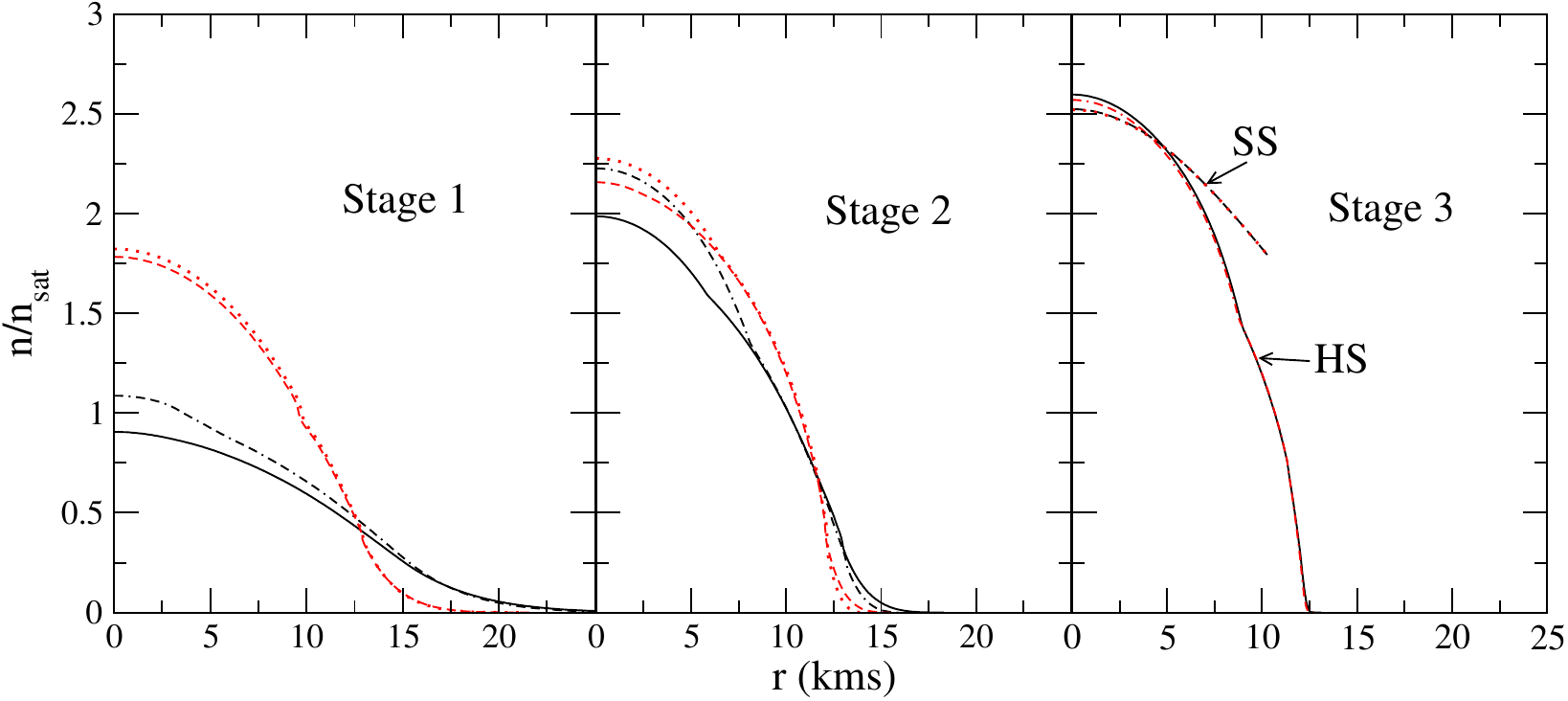}
\caption{Variation of normalized baryonic number density $n_B/n_{sat}$ with the radius of the star $r$ at different stages of fixed baryonic mass $M_B = 1.4 M_{\odot}$, for different values of $S/A$ and $Y_L$. All legends are similar to fig. \ref{fig:MvsCED}.}
\label{fig:nbvsr1.4}
\end{center}
\end{figure*}

The density profile inside the remnant of baryonic mass $1.4~M_\odot$ in different stages is shown in \cref{fig:nbvsr1.4}. It is quite evident that with the advent of time, as $S/A$ and $Y_L$ decrease, the matter density inside the star increases leading to the probability of quark matter appearance. The appearance of quark softens the matter further reducing the size and enhancing the baryonic number density near the core of the star. Naturally, in the final stage, the SS has more baryonic number density near the surface as compared to HS in the same stage due to more compactness. 
\begin{table*}
\centering
\caption{Properties of the star of maximum mass at different stages. Last two columns in stage 3 belong to SSs.}
\begin{tabular}{@{}l|clll|llll|llll@{}}
\toprule
                                                   & \multicolumn{4}{c|}{Stage 1} & \multicolumn{4}{c|}{Stage 2} & \multicolumn{4}{c}{Stage 3} \\ \midrule
                                                   \begin{tabular}[c]{@{}l@{}}$S/A$\\ ($k_B$)\end{tabular}  & 3  & 3  & 2 & 2 & 2  & 2  & 1 & 1 & 0.5  & 0.25  & 0.5 & 0.25 \\ \midrule
$Y_L$                                                  & 0.4   & 0.3   & 0.4  & 0.3  & 0.3   & 0.2   & 0.3  & 0.2  & 0     & 0     & 0    & 0    \\ \midrule
\begin{tabular}[c]{@{}l@{}}$M_{max}$\\ ($M_{\odot}$)\end{tabular}  & 2.43  & 2.45  & 2.39 & 2.41 & 2.12  & 2.08  & 2.19 & 2.14 & 2.07  & 2.07  & 2.20 & 2.20 \\ \midrule
\begin{tabular}[c]{@{}l@{}}R\\ (km)\end{tabular}   & 15.11 & 14.66 & 13.51 & 13.06 & 13.40  & 12.57 & 12.94  & 12.32 & 11.63 & 11.58 & 11.07   & 11.08  \\ \midrule
\begin{tabular}[c]{@{}l@{}}$\epsilon_c$\\ (MeV/fm$^3$)\end{tabular} & 956 & 941 & 1062 & 1063& 1044 & 1121 & 1025  & 1047 & 1170 & 1164 & 1104  & 1102\\ \bottomrule
\end{tabular}
\label{tab:1}
\end{table*}

To understand the evolution of a star let us consider an example of the star with baryonic mass $M_B=1.4~M_\odot$. The baryonic mass of the star is calculated as given in ref. \citep{1996A&A...305..871B}. Consider initially at stage 1, $S/A=2$ and  $Y_L=0.4$. Then with the EOS discussed here, the star has gravitational mass $M_G=1.36M_\odot$ and radius $R=21.68$ km with $\varepsilon_c = 280$ MeV fm$^{-3}$. Then when it passes to stage 2, with reduced $S/A=1$ and  $Y_L=0.2$, it contracts to size of radius $R=13.81$ with enhanced $\varepsilon_c = 343$ MeV fm$^{-3}$. Now with these values of $S/A$ and  $Y_L$, the quarks appear in MP at $\varepsilon = 265$ MeV fm$^{-3}$. Hence in this stage, the star contains SQM in MP up to a radius $5.01$ km from the centre. Finally, after the complete escape of neutrinos, when the star becomes stable, considering $S/A = 0.25$, it becomes a star of $M_G = 1.28M_\odot$ and $R=12.81$ km with $\varepsilon_c = 381$ MeV fm$^{-3}$. Then MP spreads from center to radius $6.71$ km. If star becomes SS at stage 3, $M_G$ reduces to $1.22 M_\odot$ and $R=10.47$ km with $\varepsilon_c = 355.7$ MeV fm$^{-3}$. Properties of this star are tabulated in table \ref{tab:2} with all values of $S/A$ and $Y_L$ at different stages.
\begin{figure*} 
\begin{center}
\includegraphics[width=15cm, keepaspectratio]{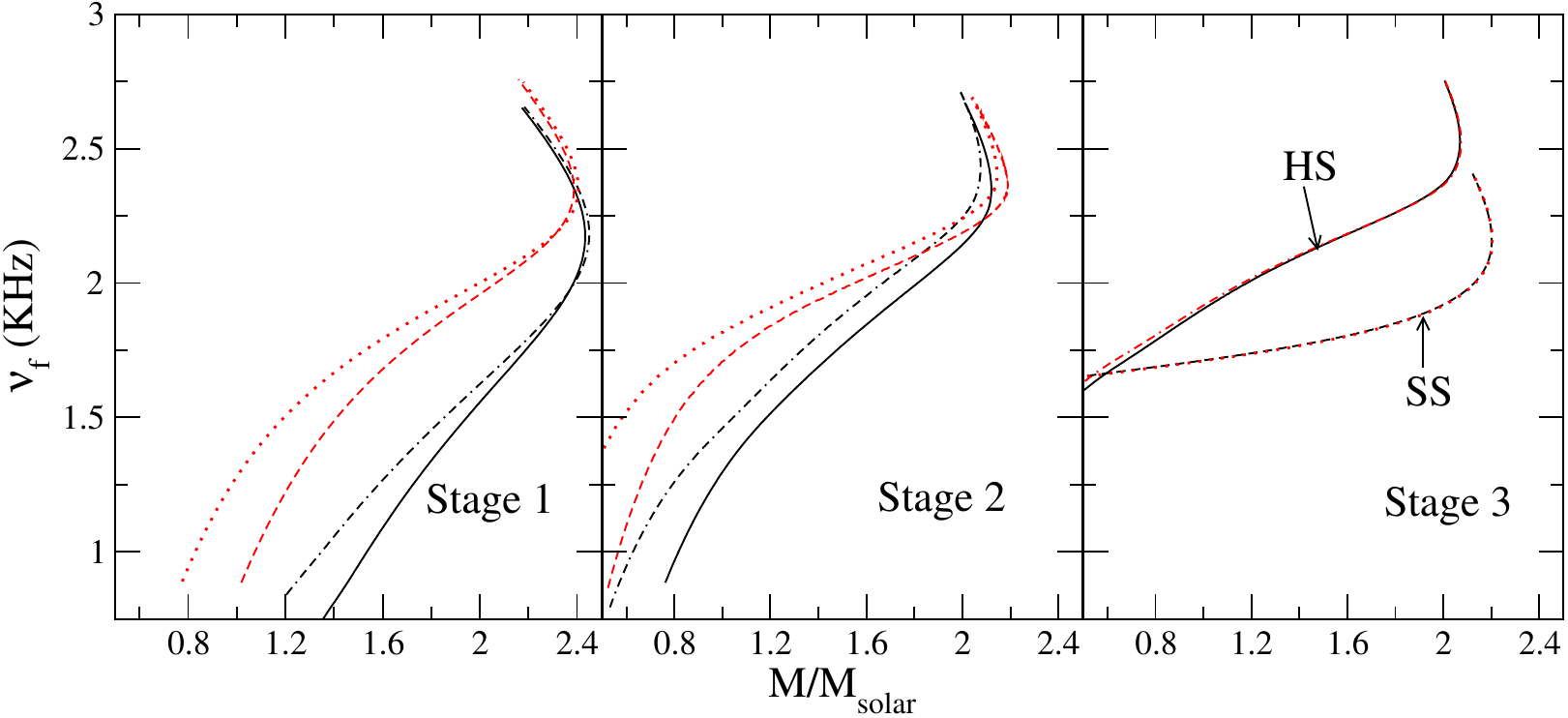}
\caption{Variation of f-mode frequency $\nu_f$ with the stellar mass in unit of solar mass $M/M_\odot$ at different stages for different values of $S/A$ and $Y_L$. All legends are similar to fig. \ref{fig:MvsCED}.}
\label{fig:fmode}
\end{center}
\end{figure*}

\begin{figure*} 
\begin{center}
\includegraphics[width=15cm, keepaspectratio]{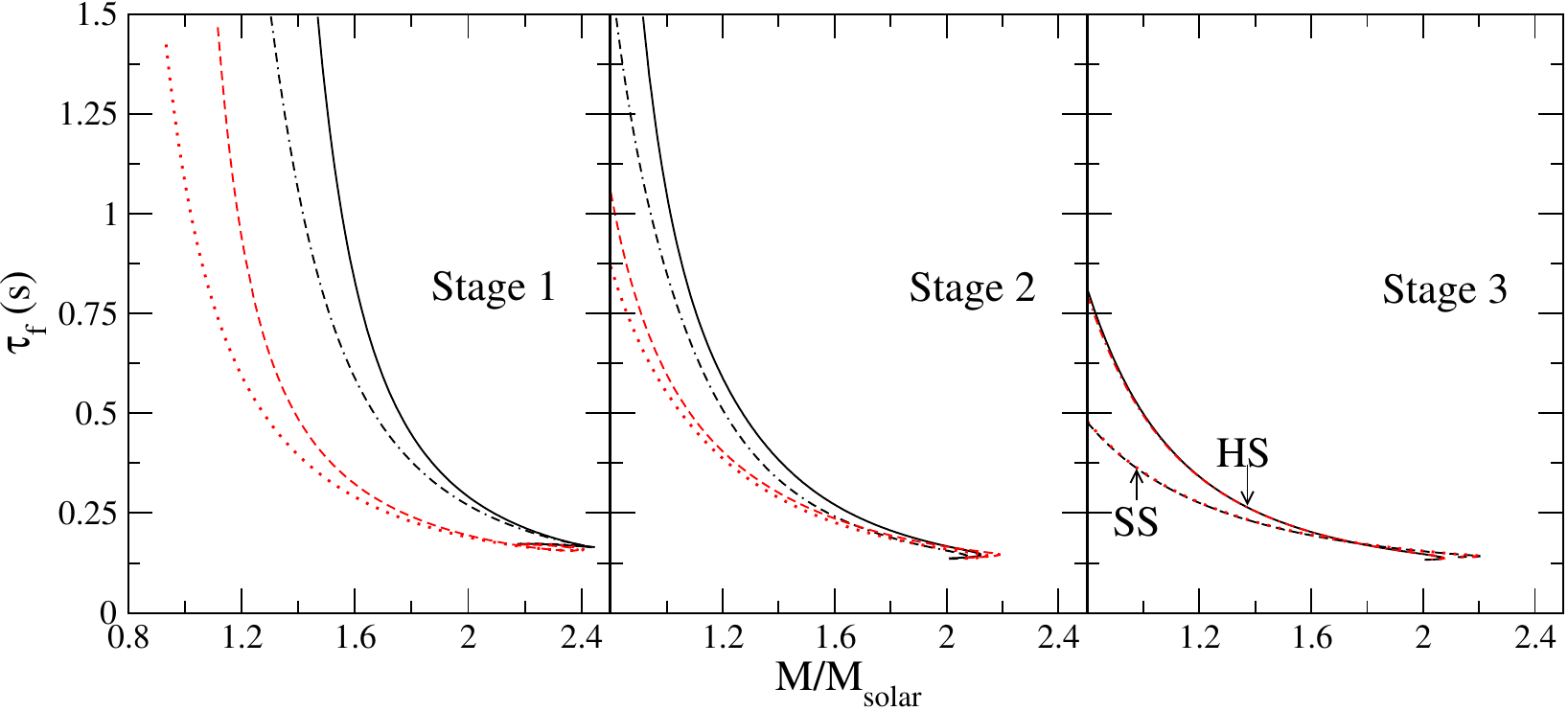}
\caption{Varitation of damping time of f-mode oscillation $\tau_f$ with stellar mass in unit of solar mass $M/M_\odot$ at different stages, for different values of $S/A$ and $Y_L$. All legends are similar to fig. \ref{fig:MvsCED}.}
\label{fig:damping}
\end{center}
\end{figure*}

\begin{figure*} 
\begin{center}
\includegraphics[width=15cm, keepaspectratio]{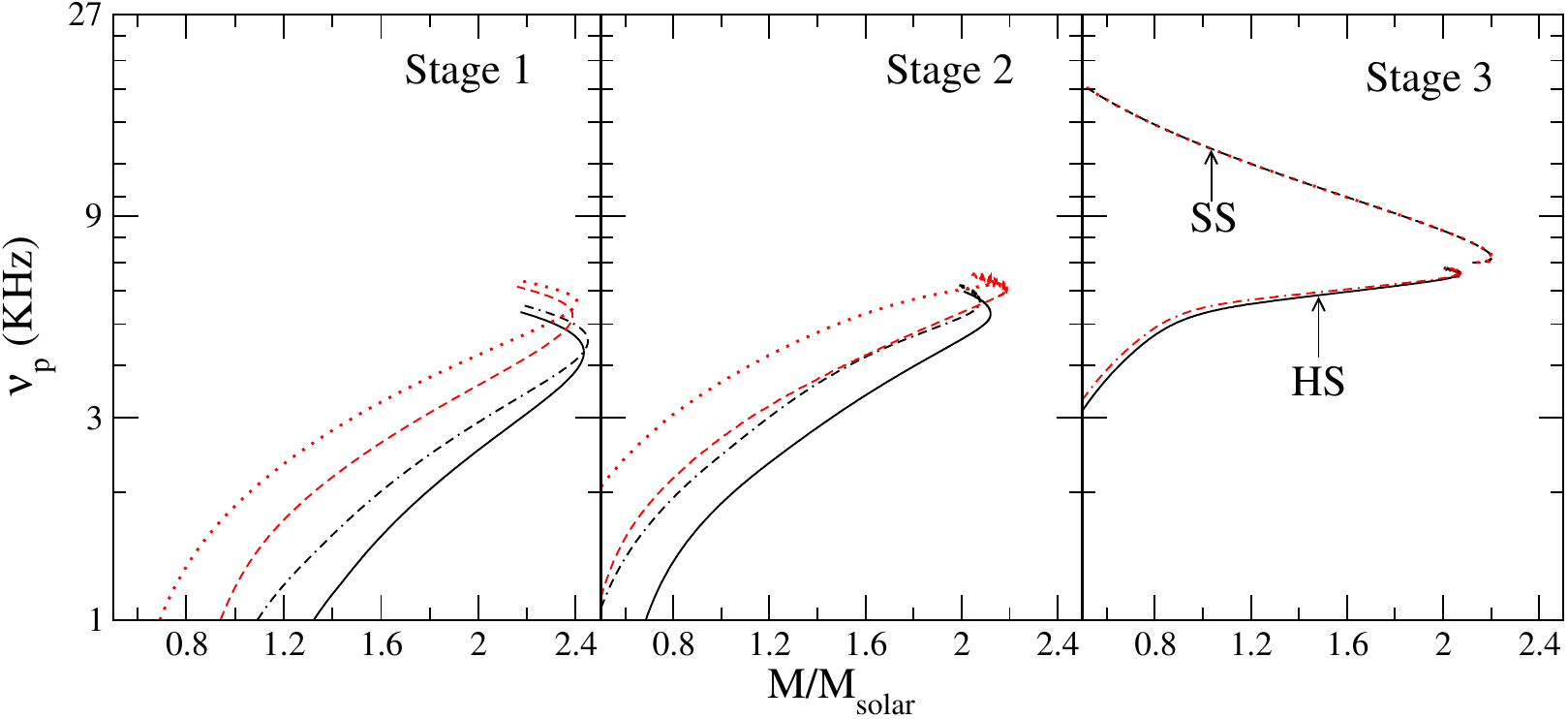}
\caption{Variation of p-mode frequency $\nu_p$ with the stellar mass in unit of solar mass $M/M_\odot$ at different stages for of $S/A$ and $Y_L$. All legends are similar to fig. \ref{fig:MvsCED}.}
\label{fig:pmode}
\end{center}
\end{figure*}

\begin{table*}
\begin{center}
\caption{Properties of the star with fixed baryonic mass $1.4M_\odot$. $\epsilon_c$, $\epsilon_s$ (MP), $r_s$ (MP), and $M_\odot$ are central energy density, starting energy density of MP, starting range of MP and gravitational mass respectively}
\begin{tabular}{ccccccccccccc}
\hline \hline
       & S/A & $Y_L$ & $M_G$ & $\epsilon_c$ & $R_{1.4}$ & $\epsilon_s$ (MP) & $r_s$ (MP) & $\nu_f$ & $\nu_p$ \\
       & ($k_B$)&  & ($M_{\odot}$) & (MeV/fm$^3$) & (km) &(MeV/fm$^3$)& (km) & (KHz)& (KHz) \\
               
\hline  & 3 & 0.4 & 1.38 & 140 & 34.9 & ... & ... &0.782 & 1.11  \\
stage 1  & 3 & 0.3 & 1.37 & 170 & 27.41 & ... & ... &1.02 & 1.52  \\
   & 2 & 0.4 & 1.36 & 280 & 21.68 & ... & ... &1.44 & 2.08  \\
   & 2 & 0.3 & 1.39 & 287 & 21.28 & ... & ... &1.66 & 2.76  \\
\\
   & 2 & 0.3 & 1.34 & 307 & 18.28 & 242 & 3.81 &1.64 & 2.70  \\
stage 2& 2 & 0.2 & 1.32 & 342 & 16.12 & 119 & 8.01 &1.74 & 3.38  \\
   & 1 & 0.3 & 1.32 & 328 & 15.50 & 319 &  3.31 &1.90 & 3.52  \\
   & 1 & 0.2 & 1.31 & 343 & 13.81 & 265  & 5.01 &1.95 & 4.33  \\
\\
stage 3   & 0.5 & 0 & 1.3 & 386 & 13.09 & 204  & 6.91 &2.05 & 5.68  \\
(HS) & 0.25 & 0 & 1.28 & 381 & 12.81 & 210 & 6.71 &2.06 & 5.78  \\
stage 3 & 0.5 & 0 & 1.22 & 356.6 & 10.46 & ... & ... &1.74 & 11.85  \\
 (SS)  & 0.25 & 0 & 1.22 & 355.7 & 10.47 & ... & ... & 1.739 & 11.84  \\
\hline
\end{tabular}
\label{tab:2}
\end{center} 
\end{table*}
Next, we explore the non-radial modes evolution of a star after its birth, considering the deconfinement of matter into quark matter at high densities. Initially, when $S/A$ and $Y_L$ are large, the star is less compact, and hence the $\nu_f$ is small. Gradually as time passes after the supernova, with the decrease of $S/A$ and $Y_L$, compactness increases and inside the star matter becomes dense which results in higher $\nu_f$. The appearance of quark matter inside the core does not make a significant change in frequency. At stage 3, if the entire star gets converted to a strange star, there is a significant drop in frequency. The evolution during different stages is depicted in fig. \ref{fig:fmode}. As for SS, the $\nu_f$ becomes almost independent of the mass of the star.  
In \cref{fig:damping}, we depict the damping time of the f-mode ($\tau_f$) for the specified frequencies. In massive stars, there is rapid dissipation of the f-mode as compared to smaller stars. During stage 1, the $\tau_f$ reaches its maximum value for $S/A=3$ and $Y_L=0.4$.  Subsequently, it exhibits a decreasing trend with decreasing values of $S/A$ and $Y_L$. Notably, $\tau_f$ is at its minimum for SSs at stage 3. This representation highlights a clear trend of decreasing $\tau_f$ with increasing compactness.
Next, we plot the $\nu_p$ in \cref{fig:pmode} with Cowling approximation. The magnitude of $ \nu_p$ is more than  $\nu_f$, otherwise, the variation of frequency with mass is almost similar to f-mode. As matter becomes denser with lower $S/A$ and $Y_L$, the $\nu_p$ increases. The frequency remains almost same with appearance of quark matter inside the core. However, for SS $\nu_p$ is very high relative to HS. This means crust plays a vital role in these non-radial oscillations.

\section{Summary and Conclusions}
Compact stars are born as compact central objects in supernova explosions. At the time of birth, it contains trapped neutrinos at high temperatures. Subsequently, it cools down. The density of a newly born star increases due accretion of matter and deleptonization. During this evolution period, with the density of the star inside, the matter composition and properties evolve as well. We discuss the possibility of SQM appearance during this evolution of central compact objects after the supernova. When the density of matter becomes a few times of nuclear saturation density, the nuclear matter may get deconfined into quark matter. The density of matter grows with time. As a result, quark content may increase with time. The final remnant of a supernova explosion can be an HS or a SS. 

In this context, we discuss here, SQM at finite temperature with the vBAG model in the context of compact star evolution to a stable star after its birth. With this matter, we constructed the HS model with Gibbs construction and discussed the properties of star during evolution. The appearance of SQM changes the internal and external properties of the star. We can indirectly detect this phase transition through the detection of neutrinos and non-radial oscillations. 

In this work, we notice the changes in frequency of f- and p-mode oscillations due to this phase transition. It is observed that a gradual increase in quark matter content inside the compact star causes a small increase in $\nu_f$. Moreover, it is observed, that if the entire star's matter gets converted into SQM, $\nu_f$ is almost independent of the mass of SS and it decreases. Due to this conversion, there can be a noticeable decrease in the frequency of the fundamental mode. This can have significant consequences, leading to a sharp peak in the f-mode frequency. That can help us to know the composition of the protoneutron star during its evolution. Although an accurate analysis of the detectability of these modes in a core-collapse supernova is lacking, with the advent of enhanced next-generation telescopes like the Cosmic Explorer and the Einstein telescope which carry about 10 times the sensitivity of Advanced LIGO, the possibility of detection of these modes increase \citep{rodriguezThreeApproachesClassification2023,2001MNRAS.320..307K,zhaoUniversalRelationsNeutron2022}. $\nu_p$ is very high and seems undetectable to next-generation GW detectors. Still, for completeness, we study its evolution during these stages. As in the case of $\nu_f$, $\nu_p$ also increases with the increase of quark content inside the star. It should be noted that $\nu_p$ is very high for low mass SS, on the other hand for the same size HS this frequency is very low. This mode of oscillation can easily distinguish SSs from neutron stars. Hence, we conclude that the future expected observation of GW because of non-radial oscillation will be able to indicate the exact nature of compact objects among many possibilities. 

\section*{ACKNOWLEDGEMENTS}
The authors acknowledge the financial support from the Science and Engineering Research Board (SERB), Department of Science and Technology, Government of India through Project No. CRG/2022/000069. The author A.K. thanks Vivek Baruah Thapa for fruitful discussions. Additionally, the authors thank the anonymous referee for constructive comments
that significantly contributed to enhancing the manuscript’s quality.

\section*{DATA AVAILABILITY}
Data sharing does not apply to this article as no data sets were generated during this study.
\\
\\
\\
\\


\bibliographystyle{mnras}
\bibliography{example} 




\appendix
\renewcommand{\theequation}{\Alph{section}\arabic{equation}}
\renewcommand{\thesection}{\Alph{section}}
\setcounter{equation}{0}

\section{Equations governing quasi-normal modes}\label{sec: appendix f mode calculation}

\subsection{General Relativistic Formalism}
The perturbations of the fluid inside the star are governed by the fluid Lagrangian displacement vector, taken as 
\begin{equation}
\begin{aligned}
    \xi^i &= \{r^{l-1}e^{-\Lambda}W(r), -r^{l-2}V(r)\partial_\theta,-r^{l-2}\sin^{-2}\theta V(r)\partial_\phi\}\mathcal{Y}_{lm}(\theta,\phi) e^{i\omega t} \label{eq:disp_vec_GR}
\end{aligned}    
\end{equation}

where $W$ and $V$ are the fluid perturbation amplitudes. There exist several approaches using which one could find the QNM frequencies, such as resonance matching \cite{1969ApJ...158....1T,chandrasekhar1991}, the method of continued fractions \cite{sotaniDensityDiscontinuityNeutron2001}, WKB \cite{kokkotasWModes1992}, etc. In this work, we employ the method of direct numerical integration \cite{lindblom1983quadrupole,detweiler1985nonradial,lujunli_ChinPhyB} to find the oscillation frequencies and damping times.
Lindblom and Detweiller \cite{lindblom1983quadrupole,detweiler1985nonradial} introduced a new fluid perturbation variable $X$, to replace $V$ in \ref{eq:disp_vec_GR}. The Lagrangian pressure variations are related to this new variable by 
\begin{equation}
    \Delta p= -r^le^{-\Phi}X\mathcal{Y}_{lm}e^{i\omega t}
\end{equation}
One can solve the perturbed Einstein equation, $\delta G^{\mu\nu}= 8\pi\delta T^{\mu\nu}$, to get all the relations between the perturbation functions inside the star. To avoid potential singularities in the eigenvalue problem, Lindblom and Detweiller pick the four independent variables to be $H_1$, $K$, $W$ and $X$. The differential equations governing these variables, and the algebraic relations of $H_0$ and $V$ are given as follows:
\begin{subequations}
    \begin{align}
        H_0 &= \left\{8\pi r^3e^{-\Phi}X- \qty[\frac{1}{2}l(l+1)(m+4\pi r^3 p)- \omega^2 r^3 e^{-2(\Lambda+\Phi)}]H_1\right.\nonumber\\
        &\left.+ [\frac{1}{2}(l+2)(l-1) r- \omega^2r^3e^{-2\Phi}\right.\nonumber\\
        &\left.-\frac{e^{2\Lambda}}{r}\qty(m+4\pi r^3p)\qty(3m-r+4\pi r^3 p) ]K\right\}\\
        &\times\qty{3m+\frac{1}{2}(l+2)(l-1) r+ 4\pi r^3 p}^{-1}\nonumber\\[3ex]
        V&= \qty{X + \frac{p'}{r}e^{\Phi-\Lambda}W - \frac{1}{2}(p+\epsilon)e^{\Phi}H_0}\times \qty{\omega^2(p+\epsilon)e^{-\Phi}}^{-1}\\[3ex]
        H_1' &= \frac{1}{r}\qty[l+1+\frac{2e^{2\Lambda}}{r}m+4\pi r^2(p-\epsilon)e^{2\Lambda}]H_1\\
        &+ \frac{e^{2\Lambda}}{r}\qty[H_0+K-16\pi(p+\epsilon)V]\\[3ex]
        K' &= \frac{H_0}{r}+ \frac{1}{2r}l(l+1)H_1- \qty[\frac{1}{r}(l+1)-\Phi']K- \frac{8\pi}{r}(p+\epsilon)e^{\Lambda}W\\[3ex]
        W' &= -\frac{1}{r}(l+1)W+ re^{\Lambda}\qty[\frac{e^{-\Phi}}{(p+\epsilon)}\dv {\epsilon}{p}X- \frac{1}{r^2}l(l+1)V+\frac{1}{2}H_0+K]\\[3ex]
        X' &= -\frac{1}{r}lX+ (p+\epsilon)e^{\Phi}\left\{\frac{1}{2}\qty[\frac{1}{r}-\Phi']H_0+ \frac{1}{2}\qty[r\omega^2 e^{-2\Phi}+ \frac{1}{2r}l(l+1)]H_1\right.\nonumber\\
        &- \frac{1}{r}\qty[4\pi(p+\epsilon)e^{\Lambda}+ \omega^2e^{\Lambda-2\Phi}-r^2\qty(\frac{e^{-\Lambda}}{r^2}\Phi')']W\\
        &\left. + \frac{1}{2}\qty[3\Phi'-\frac{1}{r}]K- \frac{1}{r^2}l(l+1)\Phi'V \right\}\nonumber    
    \end{align} \label{fluid_eq_GR}   
\end{subequations}

The system of differential and algebraic equations, eq. \eqref{fluid_eq_GR} completely describes the perturbations inside the star. The differential equations to be solved can be stored in an array $Y= \{H_{1}, K, W, X\}$. This system is singular at $r=0$ and numerically, it will blow up at values of $r$ close to $0$. Thus near the center, $Y(r)$ is approximated as $Y(r)= Y(0)+ \frac{1}{2} Y''(0)r^2+ \mathcal{O}(r^4)$ and the various terms of this approximation are given in \citep{lujunli_ChinPhyB}. 
At the surface of the star, the pressure perturbations, and thus $X$ must be $0$. To solve eq. \eqref{fluid_eq_GR}, we follow the method outlined in \citep{lindblom1983quadrupole}. We start off with 3 linearly independent solutions at the surface, and 2 linearly independent solutions at the center and integrate them to some point inside the star where they are matched. A linear combination of these solutions, with the coefficients obtained after matching gives the true values of $H_1$ and $K$ at the surface of the star. These variables and $H_0$ are the only variables defined outside the star, where the perturbation equations reduce to the Zerilli equation \cite{lujunli_ChinPhyB,zhaoUniversalRelationsNeutron2022}:
\begin{equation}
    \dv[2]{Z}{{r^*}}+ \qty[\omega^2- \mathcal{V}(r^*)]Z= 0 \label{Zerilli_eq}
\end{equation}
where $\mathcal{V}(r^*)$ is the Zerilli potential. Here $r^*$ is the tortoise coordinate, $r^*= r+ 2M\ln\qty(\frac{r}{2M}-1)$. 

The perturbed metric outside the star describes a combination of outgoing and incoming GWs, which is the general solution to the Zerilli equation. We are interested in the case of purely outgoing waves, representing the QNMs of the star. Outside the star, eq. \eqref{fluid_eq_GR} reduces to \cite{zhaoUniversalRelationsNeutron2022}: 

\begin{subequations}
    \begin{align}
        H_0 &= \left\{\qty[\omega^2r^2- (n+1)\frac{M}{r}]H_1+ \left[n\qty(1-\frac{2M}{r})- \omega^2r^2 \right.\right.\nonumber\\
        &\left.\left.+\frac{M}{r}\qty(1-\frac{3M}{r})\right]K\right\}
        \times\left\{\qty[1-\frac{2M}{r}]\qty[n+\frac{3M}{r}]\right\}^{-1}\\[3ex]
        H_1' &= -\frac{1}{r}\qty(l+1+ \frac{2M}{r}e^{-2\Phi})H_1+ \frac{1}{r}e^{-2\Phi}\qty(H_0+ K)\\[3ex]    
        K' &= \frac{H_0}{r}+ \frac{l(l+1)}{2r}H_1- \qty[\frac{l+1}{r}-\Phi']K
    \end{align}
\end{subequations}

where $n= (l-1)(l+2)/2$ and $M$ is the total mass of the star. After continuing the integration of these equations to sufficiently far away from the star ($\sim 50\omega^{-1}$), so that the solution can be approximated as a linear combination of incoming and outgoing waves, we convert these variables to the Zerilli ones using \cite{lujunli_ChinPhyB}: 

\begin{subequations}
    \begin{align}
        Z(r^*)&= \frac{\qty[k(r)-b(r)]K(r)- a(r)H_0(r)}{k(r)g(r)-h(r)}\\[3ex]
        \dv{Z(r^*)}{r^*} &= \frac{\qty[h(r)-b(r)g(r)]K(r)-a(r)g(r)H_0(r)}{h(r)-k(r)g(r)}
    \end{align}\label{zerilli_far}
\end{subequations}

Here

\begin{subequations}
    \begin{align}
        a(r)&= -\qty(nr+3M)/\qty[\omega^2r^2- \qty(n+1)M/r]\\
        b(r)&= \frac{\qty[nr(r-2M)-\omega^2 r^4+ M(r-3M)]}{(r-2M)\qty[\omega^2r^2-(n+1)M/r]}\\
        g(r)&= \frac{n(n+1)r^2+3nMr+6M^2}{r^2(nr+3M)}\\
        h(r)&= \frac{-nr^2+ 3nMr+ 3M^2}{(r-2M)(nr+3M)}\\
        k(r)&= -r^2/(r-2M) 
    \end{align}
\end{subequations}

Finally, we approximate the solution to the Zerilli equation, eq. \eqref{Zerilli_eq} as $Z(r^*)= A_-(\omega)Z_-(r^*)+ A_+(\omega)Z_+(r^*)$ where $Z_-$ represents the outgoing wave, $Z_+$ the incoming wave and $A_-$ and $A_+$ their amplitudes. At a large enough radius, 

\begin{subequations}
    \begin{align}
        Z_-&= e^{-i\omega r^*} \qty[\beta_0+\frac{\beta_1}{r}+\frac{\beta_2}{r^2}+ \mathcal{O}(r^3)]\\[3ex]
        \dv{Z_-}{r^*}&= -i\omega e^{-i\omega r^*}\qty[\beta_0+ \frac{\beta_1}{r}+ \frac{\beta_2- i\beta_1(1-2M/r)/\omega}{r^2}]
    \end{align}
\end{subequations}

Here $Z_+$ is the complex conjugate of $Z_-$ (and hence $A_+$ the complex conjugate of $A_-$) and, \cite{zhaoUniversalRelationsNeutron2022}
\begin{subequations}
    \begin{align}
        \beta_1&= \frac{-i(n+1)\beta_0}{\omega}\\
        \beta_2&= \frac{\qty[-n(n+1)+ iM\omega(3/2+ 3/n)]\beta_0}{2\omega^2}
    \end{align}
\end{subequations}

$\beta_0$ can be any complex number that represents an overall phase. By matching the solution of $Z(r_*)$ and $\dv{Z(r^*)}{r^*}$ obtained from eq. \eqref{zerilli_far} with the above equation, we can find the amplitude $A_+$ with a simple matrix inversion \citep{zhaoUniversalRelationsNeutron2022}. The frequency of the QNM corresponds to that $\omega$ which gives $A_+=0$. 

To find the QNM frequency and its damping time we first find $A_+$, which in general will be a complex number, for several real values of $\omega$ close to the original guess. We then perform a complex polynomial fitting to approximate a parabola passing through the $A_+$ points corresponding to the $\omega$ values. The root of this parabola which has a positive imaginary part is our QNM. We then take the real part of this $\omega$ and repeat the entire procedure several more times till the desired tolerance is reached. The real part of this final $\omega$ is the frequency of the QNM. The inverse of the imaginary part is the corresponding damping time \citep{PhysRevC.106.015805}. 

The numerical integration of all the relevant ODEs was done using a FORTRAN subroutine, LSODA, which automatically adjusts the step size and switches between non-stiff (Adam's) and stiff (BDF) methods. We further used a thread-safe version of this algorithm so that we could run our code in parallel on multiple threads, significantly reducing the computation time. We specified the relative tolerance to be $10^{-8}$ and the absolute tolerance to be $10^{-12}$ throughout the code and obtained satisfactory results. As with the Cowling case, the validity of our code was checked thoroughly by comparing our results with those in \citep{sotaniDensityDiscontinuityNeutron2001,2022PhRvD.106f3005K}. Moreover, we also calculated the oscillation modes and damping times using Thorne \cite{thorneNonRadialPulsationGeneralRelativistic1967} Ferrari's \cite{chandrasekhar1991} Breit-Wigner resonance fitting approach and although the damping times could not be obtained with good accuracy for all cases, the frequency values matched exactly to those obtained by the Lindblom approach \citep{lindblom1983quadrupole,detweiler1985nonradial}.

\subsection{Relativistic Cowling Approximation}\label{subsec: appendix Cowling approximation}

Since the non-radial oscillation formalism is developed in linearised gravity, the line element of a non-rotating neutron star is taken as:
\begin{equation}
 ds^2 = -e^{2\Phi(r)}dt^2 + e^{2\Lambda(r)}dr^2 + r^2d\theta^2 + r^2\sin^2{\theta}d\phi^2   
\end{equation}

The fluid Lagrangian displacement vector is assumed to be: 
\cite{2011PhRvD..83b4014S}
\begin{equation}
\begin{aligned}
\xi^i = \left(e^{-\Lambda}W,-V\partial_\theta,-V\sin^{-2}\theta\partial_\phi\right)r^{-2}\mathcal{Y}_{lm}(\theta,\phi) 
\end{aligned}
\end{equation} 
where $\mathcal{Y}_{lm}(\theta,\phi)$ are the spherical harmonics. $W$ and $V$ are perturbative fluid variables, which are functions of $r$ with a harmonic time dependence $W(t,r)=W(r)e^{i{\omega}t}$ and $V(t,r)=V(r)e^{i{\omega}t}$. The mode frequencies ($\omega$) are thus found by solving the following system of ordinary differential equations \cite{2011PhRvD..83b4014S}: 

\begin{equation}
\begin{aligned}
W' &= \dv{\epsilon}{p}\left[{\omega}^2r^2e^{\Lambda-2\Phi}V + \Phi'W\right] - l(l+1)e^{\Lambda}V\\    
V' &= 2\Phi'V-r^{-2}e^{\Lambda}W
\end{aligned}\label{eqn-mode osc}    
\end{equation}
where $\epsilon$ is the energy density and $p$ the pressure. The dash ($'$) represents a derivative with the radius. The boundary conditions at the center of the star are $W(r\to0)= Ar^{l+1}$ and $V(r\to0)=-Ar^l/l$, where $A$ is an arbitrary constant, usually taken to be $1$. 

The surface boundary condition corresponds to the fluid pressure vanishing at the surface of the star ($r= R$): 
\begin{equation}
{\omega}^2e^{\Lambda(R)-2\Phi(R)}V(R) + \frac{1}{R^2}\frac{d\Phi(r)}{dr}\bigg\vert_{r=R}W(R) = 0 \label{eqn-surf cond}    
\end{equation}

To solve these equations, we first solve the stellar structure equations for a particular central energy density, to get all the metric coefficients as functions of $r$. These coefficients are then used while solving the oscillation mode equations (eq. \eqref{eqn-mode osc}). Since we need to find $\omega$, we use the shooting method: first, a guess value of $\omega$ is used to solve the mode oscillation equations, after which we check if the surface boundary condition eq. \eqref{eqn-surf cond} is satisfied. If not, the guess is improved via the Newton-Raphson iteration scheme.  The number of radial nodes was found by counting the number of times the perturbative variables $W$ and $V$ become $0$ within $r<R$ since this condition ensures that the three-velocity of the fluid (which is the time derivative of the fluid Lagrangian displacement vector), becomes $0$ at those points. The correctness of our code was ensured when our results matched with those reported in \citep{2022PhRvD.106f3005K,thapaFrequenciesOscillationModes2023}.

\section{Equation of states}
\subsection{Equation of state of nuclear matter}\label{sec: appendix EoS NM}
Here we describe the equation of state for the nuclear matter we have used for stars in stage 1 and stage 2. The nuclear matter we consider is composed
of neutrons (n), protons(p), electrons(e), and electron neutrinos ($\nu_e$). To find the EOS of nuclear matter we consider the relativistic mean field (RMF) approach in which the nucleons interact with each other by exchange of mesons $\sigma$, $\omega$, and $\rho$. In this model the Lagrangian density is given by 
\cite{2021Univ....7..382S,2022MNRAS.517..507K}:
\begin{equation}
    {\cal L} = {\cal L}_{B} + {\cal L}_L
\end{equation}
with for nucleons ${\cal L}_{B}$ 
\begin{equation}
\begin{aligned}
    {\cal L}_{B} &= \sum_{N} \bar{\psi}_N(i\gamma_{\mu} D^{\mu} - m^{*}_N) \psi_N\\
    &- \frac{1}{2}\partial^{\mu}\sigma\partial_{\mu}\sigma - \frac{1}{2}{m_{\sigma}^2\sigma^2} - \frac{1}{4}\omega^{\mu\nu}\omega_{\mu\nu} + \frac{1}{2}m_{\omega}^2\omega^\mu\omega_\mu \\& - \frac{1}{4}\boldsymbol{\rho^{\mu\nu}}\cdot\boldsymbol{\rho_{\mu\nu}} + \frac{1}{2}m_{\rho}^2\boldsymbol{\rho^\mu}\cdot\boldsymbol{\rho_\mu},
\end{aligned}
\end{equation}
and for leptons  
\begin{equation}
\begin{aligned}
{\cal L}_L & = \sum_{l = e,\nu_e}\bar{\psi}_l\left(\gamma_{\mu}i\partial^\mu -m_l\right)\psi_l.
\end{aligned}
\end{equation}
Here $m_N$ is the mass of $N$-th particle with $N$ the nucleons, mesons, and leptons considered. The covariant derivative given by $D_{\mu} = \partial_\mu + ig_{\omega N} \omega_\mu + ig_{\rho N} \boldsymbol{\tau}_{N3} \cdot \boldsymbol{\rho}_{\mu}$. For vector mesons, field-strength tensors are as:
\begin{equation}
\begin{aligned}
    \omega_{\mu\nu} = \partial_\mu\omega_\nu - \partial_\nu\omega_\mu \\
    \rho_{\mu\nu} = \partial_\mu\rho_\nu - \partial_\nu\rho_\mu
\end{aligned}    
\end{equation}
The pressure and energy density for the nucleons are then given as
\begin{equation}
\begin{aligned}
  \epsilon_B &= \sum_{N}\frac{1}{4\pi^3}\int_{0}^{\infty}E_N(f_{N+}+f_{N-})d^3k_N\\&
  + \frac{1}{2}{m_{\sigma}^2\sigma^2} + \frac{1}{2}m_{\omega}^2\omega_0^2 + \frac{1}{2}m_{\rho}^2\rho_{03}^2 + \epsilon_L
\end{aligned}    
\end{equation}
\begin{equation}
\begin{aligned}
  p_B &= \frac{1}{3}\sum_{N}\frac{1}{4\pi^3}\int_{0}^{\infty}\frac{k_N^2}{E_N}(f_{N+}+f_{N-})d^3k_N\\&
  - \frac{1}{2}{m_{\sigma}^2\sigma^2} + \frac{1}{2}m_{\omega}^2\omega_0^2 + \frac{1}{2}m_{\rho}^2\rho_{03}^2 + p_L
\end{aligned}    
\end{equation}
where $k_N$ is the momentum of nucleon and $E_N=\sqrt{k^2+m_N^*}$ is the single particle energy for nucleons with the effective nucleon mass  
\begin{equation}
    m_N^* = m_N - g_{{\sigma}N}\sigma.
\end{equation}
Here $g_{MN}$ are the coupling parameters between nucleons and mesons ($M=\sigma,\omega~{\rm and}~\rho$).

In our present discussion, we consider the coupling parameters to be density-dependent as
\begin{equation}
\begin{aligned}
    g_{MN}(n_B) = g_{MN}(n_{sat})h_M(x)
\end{aligned}
\end{equation}
where $n_B$ is the total baryonic density and $x=n_B/n_{sat}$. For mesons $\sigma$ and $\omega$, $h_M(x)$ can be given as
\begin{equation}
    h_M(x) = \frac{a_M+b_M(x+d_M)^2}{a_M+c_M(x+d_M)^2}
\end{equation}
for $\rho$ meson this becomes as
\begin{equation}
    h_\rho = e^{-a_\rho(x-1)}
\end{equation}
The energy density and pressure of the leptons are given by 
\begin{equation}
\begin{aligned}
        \epsilon_L &= \sum_{l=e,\nu_e}\frac{2J_l+1}{2\pi^2}\int_{0}^{\infty}k^2\sqrt{k^2+m_l^2}(f_{l+}+f_{l-})dk_l\\[1.5ex]
    p_L &= \frac{1}{3}\sum_{l=e,\nu_e}\frac{2J_l+1}{2\pi^2}\int_{0}^{\infty}\frac{k^4}{\sqrt{{m_l}^2+k^2}}(f_{l+}+f_{l-})dk_l.
\end{aligned}    
\end{equation}
$2J_l+1$ is the spin degeneracy factor with the value of $2$ for electrons and 1 for neutrinos and $k_l$ denotes the momentum of each lepton. Here
\begin{equation}
f_{l\pm} = \frac{1}{1+exp[(E_l\mp\mu_l)/T]}
\end{equation}  
is the Fermi distribution function for leptons and 
with $\mu_l=\sqrt{{m_l}^2+k_{Fl}^2}$ where $k_{Fl}$ the Fermi momentum of the lepton $l$. Similarly, for nucleons the Fermi distribution function is
\begin{equation}
f_{N\pm} = \frac{1}{1+exp[(E_N\mp\mu^*_N)/T]}
\end{equation}
where for nucleons the effective chemical potential is  
\begin{equation}
    \mu_N^* = \mu_N + g_{{\omega}N}\omega_0 + g_{{\rho}N}\rho_{03}\tau_{3N} + \Sigma^r
\end{equation}
with rearrangement term ($\Sigma^r$) due to considering density-dependent parametrization is given as
\begin{equation}
\begin{aligned}
\Sigma^{r} & = \sum_{i=N} \left[ \frac{\partial g_{\omega N}}{\partial n_N}\omega_{0}n_{N} - \frac{\partial g_{\sigma N}}{\partial n_N} \sigma n_{N}^s + \frac{\partial g_{\rho N}}{\partial n_N} \rho_{03} \boldsymbol{\tau}_{b3} n_{N} \right].
\end{aligned}
\end{equation}
The scalar density ($n_{N}^s$) and vector density ($n_N$) of nucleon $N$ can be given as
\begin{equation}
\begin{aligned}
n_N &= \frac{1}{4\pi^3}\int_{0}^{\infty}(f_{N+}-f_{N-})d^3k_N \\
n_{N}^s &= \frac{1}{4\pi^3}\int_{0}^{\infty}\frac{m_N^*}{\sqrt{k^2+{m_N^*}^2}}(f_{N+}+f_{N-})d^3k_N
\end{aligned}
\end{equation}
The entropy of the system can be obtained through the following equation 
\begin{equation}
S = \frac{1}{T}\left(\epsilon + p - \sum_{i}\mu_in_i\right)
\label{eqn:entropy}
\end{equation} 

\subsection{Equation of state of SQM}\label{sec: appendix EoS SQM}
Here we consider the SQM inside the compact objects to be mainly composed of $u$, $d$, and $s$ quarks with some fractions of leptons $e$ and $\nu_e$.
The matter at this stage can be described by the Lagrangian density:
\begin{equation}
{\cal L} = {\cal L}_Q  + {\cal L}_L    
\end{equation}
where,\ ${\cal L}_Q$ and ${\cal L}_L$ are Lagrangian densities for quarks and leptons respectively. We are considering the vBAG model \cite{franzon2016effects, 2022MNRAS.513.3788K,2021PhyS...96f5302L, PhysRevD.107.063024} for the quark matter where the repulsive interactions between quarks are mediated through the repulsive field $V^\mu$, similar to the isoscalar-vector field $\omega$ in QHD \citep{glendenning2012compact,2020PhRvD.102l3007T,2021arXiv210208787B}. Here, we ignore the self-interaction term because it does not make significant changes in results \citep{2022MNRAS.513.3788K,2021PhyS...96f5302L}. Hence:

\begin{equation}
\begin{aligned}
    {\cal L}_Q &= \sum_{q = u,d,s}\left[\bar{\psi}_q\{\gamma_{\mu}\left(i\partial^\mu - g_{qV}V_\mu\right)-m_q\}\psi_q-B \right] \Theta(\bar{\psi}_q\psi_q)\\
    &- \frac{1}{4}{\left(\partial_{\mu}V_{\nu}-\partial_{\nu}V_{\mu}\right)}^2 + \frac{1}{2}{m^2_V}{V_\mu}V^{\mu}\\&
\end{aligned}
\end{equation}  
where  $m_q$ is mass of individual quark, with values $m_u=4$, $m_d=7 $ and $m_s=100$ MeV, and $m_V$ is mass of vector meson. $B$ denotes bag constant and $\Theta$ denotes the Heaviside step function that is null outside the bag and unity inside the bag. $g_{qV}$ is the coupling constant with quarks. Leptons are considered free, with ${\cal L}_{l}$ same as mentioned in previous subsection \ref{sec: appendix EoS NM}. 


Vector interaction shifts the chemical potential of free quarks as:
\begin{equation}
\mu_q = \sqrt{{m_q}^2+{k_{q}^2}} + g_{qV}V_0
\end{equation}
where $k_{q}$ is the Fermi momentum of the quark species $q$, and the mean value of $V^\mu$ in ground state is given by the relation:
\begin{equation}
g_{qV}V_0 = \left(\frac{g_{qV}}{m_V}\right)\sum_{u,d,s}{\left(\frac{g_{qV}}{m_V}\right)n_q}
\end{equation}
where the number density of individual quark flavor, $n_q$, is given by:
\begin{equation}
n_q = \frac{3}{4\pi^3}\int_{0}^{\infty}(f_{q+}-f_{q-})d^3k_q
\end{equation}
with 
\begin{equation}
f_{q\pm} = \frac{1}{1+exp[(E_q\mp\mu^*_q)/T]}.
\end{equation}   
Here, $E_q=\sqrt{{m_q}^2+k_q^2}$ and $k_q$ is momentum of quark flavor $q$. The effective chemical potential is $\mu^*_q = \mu_q - g_{qV}V_0 $. The constant $({g_{qV}}/{m_V})^2$ can be replaced by another constant $G_V$ assuming $g_{qV}$ is same for each flavor. Then the energy density and pressure of the matter are given by:
\begin{align}
    \epsilon &= \epsilon_Q + \epsilon_L -\frac{1}{2}{m_V}^2V_0^2 + B\\[1.5ex]    
    p &= p_Q + p_L + \frac{1}{2}{m_V}^2V_0^2 - B
\end{align}
where $\epsilon_L$ and $p_L$ are the energy density and pressure contribution from leptons that we mentioned previously and 
\begin{align}
    \epsilon_Q &= \sum_{u,d,s}\frac{3}{4\pi^3}\int_{0}^{\infty}\sqrt{{m_q}^2+k^2}\mu_q(f_{q+}+f_{q-})d^3k_q\\[1.5ex]
    p_Q &= \sum_{u,d,s}\frac{1}{4\pi^3}\int_{0}^{\infty}\frac{k^2}{\sqrt{{m_q}^2+k^2}}(f_{q+}+f_{q-})d^3k_q\\[1.5ex]
\end{align} 
The charge neutrality condition for the matter is as
\begin{equation}
2n_u - n_d - n_s - 3n_e = 0
\end{equation}
We consider $G_V=0.2$ fm$^2$ throughout this work.
The entropy of quark matter can be calculated by the same equation \ref{eqn:entropy}.

\bsp	
\label{lastpage}
\end{document}